\input amstex
\magnification 1200
\TagsOnRight
\def\qed{\ifhmode\unskip\nobreak\fi\ifmmode\ifinner\else
 \hskip5pt\fi\fi\hbox{\hskip5pt\vrule width4pt
 height6pt depth1.5pt\hskip1pt}}
\NoBlackBoxes \baselineskip 19 pt
\parskip 5 pt
\def\stretch {\noalign{\medskip}}
\define \bCp {\bold C^+}
\define \bCm {\bold C^-}
\define \bCpb {\overline{\bold C^+}}
\define \bCmb {\overline{\bold C^-}}
\define \ds {\displaystyle}
\define \bR {\bold R}
\define \bm {\bmatrix}
\define \endbm {\endbmatrix}

\centerline {\bf SMALL-ENERGY ANALYSIS FOR}
\centerline {\bf  THE SELFADJOINT MATRIX SCHR\"ODINGER OPERATOR}
\centerline {\bf ON THE HALF LINE}

\vskip 10 pt
\centerline {Tuncay Aktosun}
\vskip -8 pt
\centerline {Department of Mathematics}
\vskip -8 pt
\centerline {University of Texas at Arlington}
\vskip -8 pt
\centerline {Arlington, TX 76019-0408, USA}
\vskip -8 pt
\centerline {aktosun\@uta.edu}

\centerline {Martin Klaus}
\vskip -8 pt
\centerline {Department of Mathematics}
\vskip -8 pt
\centerline {Virginia Tech}
\vskip -8 pt
\centerline {Blacksburg, VA 24061, USA}
\vskip -8 pt
\centerline {mklaus\@math.vt.edu}

\centerline {Ricardo Weder\plainfootnote{$^\dagger$}
{Fellow Sistema Nacional de Investigadores}}
\vskip -8 pt
\centerline {Instituto de Investigaciones en
Matem\'aticas Aplicadas y en Sistemas}
\vskip -8 pt
\centerline {Universidad Nacional Aut\'onoma de M\'exico}
\vskip -8 pt
\centerline {Apartado Postal 20-726, IIMAS-UNAM, M\'exico DF 01000, M\'exico}
\vskip -8 pt \centerline {weder\@unam.mx}

\noindent {\bf Abstract}: The matrix Schr\"odinger equation
with a selfadjoint matrix potential is considered on the half
line with the most general selfadjoint boundary condition at
the origin. When the matrix potential is integrable and has a
first moment, it is shown that the corresponding scattering
matrix is continuous at zero energy. An explicit formula is
provided for the scattering matrix at zero energy. The
small-energy asymptotics are established also for the related
Jost matrix, its inverse, and various other quantities relevant
to the corresponding direct and inverse scattering problems.

\vskip -8 pt
\par\noindent {\bf Short title:} Half-line matrix Schr\"odinger
equation

\newpage

\noindent {\bf 1. INTRODUCTION}
\vskip 3 pt

Consider the matrix Schr\"odinger equation on the half line
$$-\psi''+V(x)\,\psi=k^2\psi,\qquad x\in(0,+\infty),\tag 1.1$$
where the prime denotes the derivative with respect to
the spatial coordinate $x$ and
the potential $V$ is a $n\times n$ matrix-valued function
belonging to class $L^1_1(\bR^+)$ with $\bR^+:=(0,+\infty).$
Note that $V\in L^1_j(\bold I)$ means that
each entry of the matrix $V$ is Lebesgue measurable on the interval
$\bold I$ and
$$\int_{\bold I}
dx\,(1+|x|)^j||V(x)||<+\infty,$$ where $||V(x)||$ denotes a
matrix norm. Clearly, a matrix-valued function belongs to
$L^1_1(\bR^+)$ if and only if each entry of that matrix belongs
to $L^1_1(\bR^+).$ Note that $V$ is not assumed to be real
valued, but we impose the condition that it is selfadjoint,
i.e.
$$V=V^\dagger,\tag 1.2$$
where the dagger denotes the adjoint (complex
conjugate and matrix transpose).
Without loss of any generality we can view
the wavefunction $\psi(k,x)$ appearing in (1.1)
either as a vector-valued
function with $n$ components
or as an $n\times p$ matrix-valued function
for some $p$ with $1\le p\le n.$

We are interested in studying (1.1) with a selfadjoint potential $V$
in $L^1_1(\bR^+)$ under the most general
selfadjoint boundary condition at $x=0.$ This is the generalization of
the scalar version (with $n=1$) of the corresponding problem, where the
most general selfadjoint boundary condition at $x=0$ can be stated as [5,8,16,30]
$$(\cos \theta)\,\psi(0)+(\sin\theta)\,\psi'(0)=0,\tag 1.3$$
where the parameter $\theta$ takes values in the interval $(0,\pi].$ The special choice
$\theta=\pi$ corresponds to the Dirichlet boundary condition, and the choice
$\theta=\pi/2$ corresponds to the Neumann boundary condition.

A formulation of the
most general selfadjoint
boundary condition at $x=0$ for (1.1) was stated in [23,24] as
$$A_1\,\psi(0)+B_1\,\psi'(0)=0,\tag 1.4$$
such that the constant $n\times n$ matrices
$A_1$ and $B_1$ satisfy
$$A_1B_1^\dagger=B_1 A_1^\dagger,\tag 1.5$$
$$\text{rank}\bm A_1&B_1\endbm=n,\tag 1.6$$
i.e. $A_1B_1^\dagger$ is selfadjoint and
the $n\times 2n$ matrix $\bm A_1&B_1\endbm$ has rank $n.$

Another formulation of the most general selfadjoint
boundary condition at $x=0$ for (1.1) was stated in [18-20] in terms of
a constant $n\times n$ unitary matrix $U_2$ as
$$-B_2^\dagger\psi(0)+A_2^\dagger\psi'(0)=0,\tag 1.7$$
where the auxiliary constant $n\times n$ matrices $A_2$ and $B_2$
are given by
$$A_2:=\ds\frac12\left(U_2+I_n\right),\quad
B_2:=\ds\frac i2\left(U_2-I_n\right),\tag 1.8$$
with $I_n$ denoting the $n\times n$ identity matrix. It can
directly be verified that the matrices $A_2$ and $B_2$ satisfy
$$A_2^\dagger A_2=A_2 A_2^\dagger,\quad B_2^\dagger B_2=B_2 B_2^\dagger,
\quad A_2 B_2^\dagger=B_2 A_2^\dagger,\quad A_2^\dagger B_2=B_2^\dagger A_2,\tag 1.9$$
$$A_2 A_2^\dagger+B_2 B_2^\dagger=I_n,\quad
A_2+iB_2=I_n,\quad A_2-iB_2=U_2.
\tag 1.10$$

We ourselves find it convenient to state the most general
selfadjoint boundary condition at $x=0$ for (1.1) in terms of
constant $n\times n$ matrices $A_3$ and $B_3$ such that
$$-B_3^\dagger\psi(0)+A_3^\dagger\psi'(0)=0,\tag 1.11$$
$$-B_3^\dagger A_3+A_3^\dagger B_3=0,\tag 1.12$$
$$A_3^\dagger A_3+B_3^\dagger B_3>0,\tag 1.13$$
i.e. $A_3^\dagger B_3$ is selfadjoint and
the selfadjoint matrix $(A_3^\dagger
A_3+B_3^\dagger B_3)$ is positive. Note that (1.13)
implies the existence of a unique positive matrix $E_3$ defined
as
$$E_3:=(A_3^\dagger A_3+B_3^\dagger B_3)^{1/2},\tag 1.14$$
such that $E_3$ is selfadjoint and invertible, and hence
$$E_3=E_3^\dagger,\qquad (E_3^\dagger)^{-1}(A_3^\dagger A_3
+B_3^\dagger B_3)E_3^{-1}=I_n.\tag 1.15$$
Let us define the matrices $C_3$ and $H_3$ as follows
$$C_3:=\bm B_3&A_3\\
\stretch
A_3&-B_3\endbm,\quad H_3:=C_3 \bm E_3^{-1}&0
\\
0&E_3^{-1}\endbm.$$
With the help of (1.14) and (1.15), it can be checked that
$H_3^\dagger H_3=I_{2n}$ and hence $H_3$ is unitary. Thus,
we must have $H_3 H_3^\dagger=I_{2n},$ yielding
$$A_3E_3^{-2}A_3^\dagger+B_3E_3^{-2}B_3^\dagger=I_{2n},
\quad B_3E_3^{-2}A_3^\dagger-A_3E_3^{-2}B_3^\dagger=0.
\tag 1.16$$

We will mainly be working with the formulation given in
(1.11)-(1.13) and hence later we will drop the subscripts in
$A_3,$ $B_3,$ $E_3$ and simply write $A,$ $B,$ $E$ if there is
no confusion.

Let us note that one can multiply the boundary conditions
stated in (1.4), (1.7), and (1.11) on the left by
an invertible matrix $D$ without changing the most general
selfadjoint boundary condition at $x=0.$ For example, for
(1.11), by dropping the subscript $3,$ that left multiplication can be
described via the
transformation
$$(A,B)\mapsto (\tilde A,\tilde B):=(AD^\dagger,BD^\dagger),\tag 1.17$$
for which we have
$$(-B^\dagger A+A^\dagger B)\mapsto -D^{-1}(-\tilde B^\dagger
\tilde A+\tilde A^\dagger \tilde B)(D^\dagger)^{-1},$$
$$(A^\dagger A+B^\dagger B)\mapsto D^{-1}(\tilde A^\dagger \tilde A
+\tilde B^\dagger \tilde B)(D^\dagger)^{-1},$$
and
hence (1.11)-(1.13) hold with $(\tilde A,\tilde B)$ appearing
instead of $(A,B).$ Thus, the transformed pair $(\tilde
A,\tilde B)$ can be used instead of $(A,B)$ in the formulation
of the selfadjoint boundary condition.

Our primary goal is to establish, under the most general
selfadjoint boundary condition at $x=0,$ the small-$k$
asymptotics of various quantities related to (1.1) such as
scattering solutions, the Jost matrix, the inverse of the Jost
matrix, and the scattering matrix. The small-$k$ analysis for
(1.1) has been lacking in the literature even though the
relevant results are crucial in the study of the corresponding
direct and inverse scattering problems. The direct scattering
problem for (1.1) is to determine the scattering matrix and the
bound-state information when the matrix potential $V$ and the
selfadjoint boundary condition are known. On the other hand,
the inverse scattering problem is to recover the potential and
the boundary condition from an appropriate set of scattering
data. In some sense, our paper can be considered as a
complement to the study by Agranovich and Marchenko [1], where
the inverse scattering problem is analyzed only under the
Dirichlet boundary condition but with attention to the behavior
at $k=0.$ Our study can also be considered as a complement to
the study by Harmer [18-20] where the most general selfadjoint
boundary condition (1.7) is used to investigate the inverse
problem for (1.1) but the small-$k$ analysis is omitted. We
refer the reader to [2,22] for similar small-$k$ analyses for
the scalar radial Schr\"odinger equation, to [3,21] for the
scalar full-line Schr\"odinger equation, to [4] for the matrix
full-line Schr\"odinger equation, and to [5] for the radial
Schr\"odinger equation with the most general selfadjoint
boundary condition at the origin.

Let us look at the definition of the Jost matrix $J(k)$ given in
(4.3). When $V$ is selfadjoint and belongs to
$L_1^1(\bR^+),$ it is already known [1,4] that
the right hand side in (4.3) is continuous
in $k\in\bCpb,$ where we use $\bold C$ for the
complex plane, $\bCp$ for
the upper half complex plane, and
$\bCpb:=\bCp\cup\bR.$ Thus, $J(0)$ exists.
Let us also look at the definition of the scattering
matrix $S(k)$ given in (4.6) in terms of the Jost matrix $J(k).$
In case $J(0)$ is invertible, it is clear from (4.6) that $S(0)=-I_n.$
However, if $J(0)$ is not invertible, it is unclear
whether $S(k)$ is continuous at $k=0$ and what the value of
$S(0)$ is in case the continuity at $k=0$ is assured. Our paper
mainly concentrates on the case when $J(0)^{-1}$ does not
exist. We prove that $S(k)$ is indeed continuous at
$k=0$ and we determine the value of $S(0),$ which
in general is different from $-I_n.$
In case $J(0)$ is invertible,
our results reduce to the easy case with $S(0)=-I_n.$

Let us note that $J(0)$ is not invertible if and only if
the determinant $\det[J(0)]$ is zero. In the scalar
case (i.e. when $n=1$) this is the analog of $F_\theta(0)=0,$
where $F_\theta(k)$ is the Jost function appearing in
(4.1). The case $F_\theta(0)=0$ is known as the ``exceptional case,"
and the case $F_\theta(0)\ne 0$ is known as the ``generic case."
Hence, in our paper we concentrate on the ``exceptional case"
for (1.1), namely the case when $J(0)$ is not invertible.
In the ``generic case" it is already known and easy to see
that $S(k)$ is continuous at
$k=0$ and $S(0)=-I_n.$ In the exceptional case, by
expressing $J(k)$ as in (6.17) in terms of a related
matrix $\Cal Z(k),$ and by writing the scattering matrix
$S(k)$ as in (6.19) in terms of $\Cal Z(-k)$ and $\Cal Z(k)^{-1},$
we are able to prove the continuity of $S(k)$ at $k=0$
and evaluate $S(0).$

We remind the reader that the continuity of the
scattering matrix in the exceptional case is not
an easy matter. For example, in the full-line scalar case,
Deift and Trubowitz [9] stated that
the characterization of the scattering data
given by Faddeev [12] might not hold and in fact
even the continuity
of the scattering matrix was not clear when the real-valued
potential belonged to
$L_1^1(\bR)$ and they introduced the
stronger condition that the potential belonged to
$L_2^1(\bR).$ The proof of the continuity of the
scattering matrix when the
potential belongs to
$L_1^1(\bR)$ was given later.
For further details we refer the reader to
[3,21] and the references therein.

The matrix Schr\"odinger equation (1.1)
has direct relevance to scattering in quantum mechanics involving
particles of internal structures as spins,
scattering on
graphs [6,7,11,14,15,17,25-28], and quantum wires [23,24].
For example, the problem
under study describes $n$ connected very thin
quantum wires forming a one-vertex graph with open ends.
A linear boundary condition is imposed at the vertex and the behavior
on each wire is governed by the Schr\"odinger operator.
The problem has physical relevance to designing elementary gates
in quantum computing and nanotubes for microscopic electronic devices,
where, for example, strings of atoms may form a star-shaped graph.
For the details we refer the reader
to [23,24] and the references therein.

Our paper is organized as follows.
In Section~2 we show that
the three selfadjoint boundary
condition formulations given in Section~1 are
equivalent.
In Section~3 we introduce various $n\times n$ matrix
solutions to (1.1) and state their properties relevant
to the small-$k$ analysis of (1.1).
In Section~4 we introduce the Jost matrix $J(k)$
and the scattering matrix $S(k).$ In Section~5
we obtain various results that are crucial in determining
the small-$k$ asymptotics of the Jost matrix,
its inverse, and the scattering matrix.
In Section~6 we provide
the small-$k$ asymptotics for $J(k),$ $J(k)^{-1},$ and $S(k),$
and we prove that $S(k)$ is continuous at $k=0.$
Finally, in Section~7 we provide some examples to
illustrate the theory presented.

\vskip 10 pt \noindent {\bf 2. EQUIVALENCE OF BOUNDARY
CONDITION FORMULATIONS} \vskip 3 pt

In Section 1 we have stated the three formulations of the most
general selfadjoint boundary conditions at $x=0$ for (1.1):

\item{(a)} The formulation (1.4)-(1.6) stated in [23,24].

\item{(b)} The formulation (1.7) and (1.8) stated in [18-20].
\item{(c)} Our own formulation stated as (1.11)-(1.13).

In this section we show that those three formulations are equivalent.

\noindent {\bf Theorem 2.1} {\it The three formulations (a),
(b), and (c) of the most general selfadjoint boundary
conditions at $x=0$ for (1.1) are all equivalent.}

\noindent PROOF: With the help
of (1.17), we can relate (1.11)-(1.14) to (1.7) and (1.8) by
letting
$$U_2=(A_3-iB_3)E_3^{-2}(A_3^\dagger-iB_3^\dagger),$$
and we can verify (1.7) and (1.8) with the help of (1.15) and (1.16).
Hence, we have shown that (c) implies (b).
Next, we will show that (b)
implies (a). Let $A_1=-B_2^\dagger$ and $B_1=A_2^\dagger.$ Then
(1.7) implies (1.4). Furthermore, the last equality in (1.9)
yields (1.5). Let
$$C_2:=\bm B_2&A_2\\
\stretch
A_2&-B_2\endbm.$$
We then get
$$C_2^\dagger C_2=\bm
B_2^\dagger B_2+A_2^\dagger A_2&B_2^\dagger A_2-A_2^\dagger B_2\\
\stretch
A_2^\dagger B_2-B_2^\dagger A_2&A_2^\dagger A_2+B_2^\dagger
B_2\endbm=\bm I_n&0\\
\stretch
0&I_n\endbm,$$
where we have used (1.9) and (1.10).
Thus, $C_2$ is unitary and has rank $2n.$ As a result the
block matrix $\bm B_2^\dagger&A_2^\dagger\endbm$ has rank $n.$
That matrix is in fact equal to $\bm -A_1&B_1\endbm,$ and changing the
signs in the first $n$ columns does not affect its rank. Thus,
$\bm A_1&B_1\endbm$ has rank $n$ and (1.6) is satisfied. Hence,
we have shown that (b) implies (a). Finally, let us show that
(a) implies (c). Let $B_3=-A_1^\dagger$ and $A_3=B_1^\dagger.$
Then, (1.4) implies (1.11), and (1.5) yields (1.12). Note that
$$
A_3^\dagger A_3+B_3^\dagger B_3=\bm B_1&A_1\endbm \bm
B_1^\dagger\\
\stretch
A_1^\dagger\endbm,\tag 2.1$$
and we need to show that the
matrix product on the right in (2.1) is positive. This is
indeed the case because that matrix product is itself a
selfadjoint matrix and zero cannot be one of its eigenvalues.
Otherwise, we would have a nonzero eigenvector $v$ with
the zero eigenvalue, implying $$0=\langle v,\bm B_1&A_1\endbm
\bm
B_1^\dagger\\
\stretch A_1^\dagger\endbm v\rangle= \langle \bm
B_1^\dagger\\
\stretch A_1^\dagger\endbm v,\bm
B_1^\dagger\\
\stretch A_1^\dagger\endbm v\rangle,\tag 2.2$$ with
$\langle\cdot,\cdot\rangle$ denoting the standard scalar
product. However, (2.2) would then imply that $\bm
B_1^\dagger\\
A_1^\dagger\endbm v=0$ and hence the kernel of the matrix $\bm
B_1^\dagger\\
A_1^\dagger\endbm$ would contain the nonzero vector $v.$
Consequently, the nullity of $\bm
B_1^\dagger\\
A_1^\dagger\endbm$ would be at least 1.
Since the nullity and the rank must add up to
$n,$ the rank would have to be strictly less than $n,$
violating the fact that the rank of that matrix is exactly $n$
because of (1.6). Thus, (a)
implies (c). \qed

In the following proposition we state
a fourth equivalent formulation of the most general
selfadjoint boundary condition at $x=0$ for (1.1).

\noindent {\bf Proposition 2.2} {\it The three formulations (a),
(b), and (c) of the most general selfadjoint boundary
condition are also equivalent to the formulation in terms of
two constant $n\times n$ matrices $A_4$ and $B_4$ as}
$$-B_4^\dagger\psi(0)+A_4^\dagger\psi'(0)=0,\tag 2.3$$
{\it such that the matrix $C_4$ is unitary, where we have
defined}
$$C_4:=\bm B_4&A_4\\
\stretch
A_4&-B_4\endbm.\tag 2.4$$

\noindent PROOF: Because of Theorem~2.1, it is sufficient
to prove the equivalence of (2.3)-(2.4) with
(1.11)-(1.13). Suppose that (2.3)-(2.4) hold. Then, letting
$A_3=A_4$ and $B_3=B_4,$ we get
(1.11) and we obtain (1.12) and (1.13) from the unitarity of $C_4.$
Conversely, suppose (1.11)-(1.13) hold. By letting
$A_4=A_3 E_3^{-1}$ and $B_4=B_3 E_3^{-1},$ we observe that
(1.12) and (1.15) yield $C_4^\dagger C_4=I_{2n},$
implying the unitarity of $C_4.$ \qed

We have seen in (1.7) and (1.8) that the most general
selfadjoint boundary condition for (1.1) can be stated in terms
of a unitary matrix. There are certainly other choices for such
a unitary matrix besides $U_2$ appearing in (1.8). For example,
in terms of a unitary matrix $U_5,$ instead of (1.7) and (1.8)
we can use
$$-B_5^\dagger\psi(0)+A_5^\dagger\psi'(0)=0,\tag 2.5$$
where the auxiliary constant $n\times n$ matrices $A_5$ and
$B_5$ are given by
$$A_5:=\ds\frac i2\left(U_5-U_5^\dagger\right),
\quad B_5:=\ds\frac 12\left(U_5+U_5^\dagger\right). \tag 2.6$$
As seen from (2.6) we can simultaneously diagonalize $U_5$ into
the form
$$U_5=\text{diag}\{e^{i\theta_1},e^{i\theta_2},\dots,e^{i\theta_n}\},\tag 2.7$$
for some real-valued parameters $\theta_j.$ Then, the boundary
condition given in (2.5) is separated into the $n$ conditions
given by
$$\left(\cos \theta_j\right) \psi_j(0)+\left(\sin \theta_j\right)\psi_j'(0)=0,
\qquad j=1,\dots,n,$$
where $\psi_j$ denotes the $j$th
column of the $n\times n$ matrix solution $\psi.$ Similarly, for the
choice (1.8) for $(A_2,B_2)$ in terms of a unitary matrix
$U_2,$ by diagonalizing $U_2$ as in (2.7), we can express (1.7)
as $n$ separate boundary conditions given by
$$\left[\sin( \theta_j/2)\right] \psi_j(0)+\left[\cos( \theta_j/2)\right]\psi_j'(0)=0,
\qquad j=1,\dots,n.$$

\vskip 10 pt
\noindent {\bf 3. PRELIMINARIES}
\vskip 3 pt

In this section we introduce certain $n\times n$ matrix
solutions to (1.1) and
state their properties that will be useful later on. We state the
results without proofs and refer the reader to
the appropriate references such as [1,4] for details.
Let us recall that we use the boundary conditions
stated in (1.11)-(1.13) without the subscript $3.$
When $V$ is selfadjoint and belongs
to $L^1_1(\bold R^+),$ the matrix Schr\"odinger
equation (1.1) has various $n\times n$ matrix solutions satisfying certain
initial conditions or certain asymptotic
conditions, and the existence of such solutions
are already known.

The Jost solution to
(1.1) is the $n\times n$ matrix solution
satisfying, for $k\in\bCpb\setminus\{0\},$ the asymptotics
$$f(k,x)=e^{ikx}[I_n+o(1)],\quad
f'(k,x)=ik\,e^{ikx}[I_n+o(1)],\qquad x\to+\infty.\tag 3.1$$ It
satisfies the integral equation
$$f(k,x)=e^{ikx}I_n+\ds\frac{1}{k}\int_x^\infty dy\,\sin k(y-x)\,V(y)\,f(k,y),$$
and it is known [1,4] that
$f(k,x)$ and $f'(k,x)$ are analytic in $k\in\bCp$
and continuous in $k\in\bCpb$ for each fixed $x.$
The zero-energy Jost solution $f(0,x)$ satisfies
$$f(0,x)=I_n+\int_x^\infty dy\, (y-x)\,V(y)\,f(0,y),$$
and it is known [1,4] that $f(0,x)$ is a bounded solution
to the $n\times n$ matrix-valued zero-energy Schr\"odinger equation
$$-\psi''+V(x)\,\psi=0,\qquad x\in(0,+\infty),\tag 3.2$$
satisfying
$$f(0,x)=I_n+o(1),\quad f'(0,x)=o(1/x),\qquad x\to+\infty.\tag 3.3$$
It is also known [1,4] that (3.2) has an $n\times n$ matrix
solution $g(0,x)$ satisfying
$$g(0,x)=x[I_n+o(1)],\quad g'(0,x)=I_n+o(1),\qquad x\to+\infty.\tag 3.4$$
Thus, the $2n$ columns of $f(0,x)$ and $g(0,x)$
form a fundamental set of solutions to (3.2), and any vector
solution $\phi(x)$ to (3.2) can be expressed as
$$\phi(x)=f(0,x)\,\xi+g(0,x)\,\eta,\qquad x\in(0,+\infty),\tag 3.5$$
where the constant vectors $\xi$ and $\eta$ in $\bold C^n$
are uniquely determined by $\phi(x).$
We see from (3.4) and (3.5) that any solution to (3.2) that behaves as $o(x)$ as
$x\to+\infty$ must be a bounded solution.

There are various $n\times n$ matrix solutions to (1.1) defined
via specifying some constant initial conditions at a finite
$x$-value. As a result, such solutions are analytic in $k$ in
the entire complex plane for each fixed $x.$ Because of their
analyticity such solutions are usually called ``regular"
solutions. The $n\times n$ regular solution $\varphi(k,x)$
satisfies the initial conditions
$$\varphi(k,0)=A,\quad \varphi'(k,0)=B,\tag 3.6$$
where $A$ and $B$ are the matrices appearing in (1.11). It
satisfies the integral relation
$$\varphi(k,x)=A\,\cos kx+B\,\ds\frac{\sin kx}{k}+\ds\frac{1}{k}\int_0^x dy\,\sin k(x-y)\,V(y)\,
\varphi(k,y).\tag 3.7$$

Let us define two additional regular $n\times n$ matrix-valued
solutions to (1.1), namely $C(k,x)$ and $S(k,x)$ with initial conditions
at $x=a,$ at which the matrix $f(0,a)$ is invertible. The
existence of such an $a$-value is assured by the fact that
$f(0,x)=I_n+o(1)$ as $x\to+\infty$ and hence $f(0,x)$ is
invertible at least for large $x$-values. In fact, if
$f(0,a)^{-1}$ exists, then we must have the existence
of $f(k,a)^{-1}$
in the vicinity of $k=0$ in $\bCpb.$ This is because
for each fixed $x$-value it is known
[4] that
$f(k,x)$ is a continuous function of $k\in
\bCpb.$ Hence,
$\det[f(k,a)]$ is a continuous function of $k$ and
if it is nonzero at $k=0$ it must be nonzero
in the vicinity of $k=0.$ Thus, we conclude that
$$f(k,a)=f(0,a)+o(1),\quad f(k,a)^{-1}=f(0,a)^{-1}+o(1),
\qquad k\to 0 \text{ in }\bCpb.\tag 3.8$$

The cosine-like
solution $C(k,x)$ satisfies the initial conditions
$$C(k,a)=I_n, \quad C'(k,a)=0,\tag 3.9$$
and the sine-like solution $S(k,x)$ satisfies
$$S(k,a)=0, \quad S'(k,a)=I_n.\tag 3.10$$
Thus, we have the integral representations
$$C(k,x)=I_n\,\cos k(x-a)+\ds\frac{1}{k}\int_a^x dy\,\sin k(x-y)\,V(y)\,C(k,y),\tag 3.11$$
$$S(k,x)=I_n\,\ds\frac{\sin k(x-a)}{k}+\ds\frac{1}{k}\int_a^x dy\,\sin k(x-y)\,V(y)\,S(k,y).\tag 3.12$$
Note that we suppress the dependence on $a$ in our notation for
such solutions.

We define another $n\times n$ regular solution to (1.1),
$\omega(k,x),$ which satisfies the initial conditions
$$\omega(k,a)=f(0,a),\quad \omega'(k,a)=f'(0,a).\tag 3.13$$
Again we suppress the dependence on $a$ in our notation for $\omega(k,x).$
Note that
$$\omega(0,x)=f(0,x),\qquad x\in\bR^+,\tag 3.14$$
because both sides satisfy (1.1) when $k=0$ and they both satisfy the same initial
conditions at $x=a$ given in (3.13).
It is seen from (3.9), (3.10), and (3.13) that
$$\omega(k,x)=C(k,x)\,f(0,a)+S(k,x)\,f'(0,a),\tag 3.15$$
where $f(k,x)$ is the Jost solution appearing in (3.1).

Let us note that our regular solutions satisfy for $k\in\bold C$
$$\varphi(-k,x)=\varphi(k,x),\quad
C(-k,x)=C(k,x),\quad
S(-k,x)=S(k,x),\quad
\omega(-k,x)=\omega(k,x).\tag 3.16$$
This is because $k$ appears as $k^2$ in (1.1) and
the initial values of those solutions
are independent of $k,$ as seen from
(3.6), (3.9), (3.10), and (3.13).

Associated with (1.1) we have the adjoint equation
$$-\psi^{\dagger \prime\prime}+\psi^\dagger\,V(x)=(k^*)^2\psi^\dagger,
\qquad x\in(0,+\infty),\tag 3.17$$ where we have used (1.2) and
an asterisk denotes complex conjugation. Note that if
$\psi(k,x)$ is any solution to (1.1), then $\psi(\pm
k^*,x)^\dagger$ is a solution to (3.17). Let us also
add that if $\psi(k,x)$ has an analytic extension from $k\in\bR$ to
$k\in\bCp,$ then $\psi(-k,x)^\dagger$ has also
an analytic extension from $k\in\bR$ to
$k\in\bCp,$ and in fact that extension becomes equal to
$\psi(-k^*,x)^\dagger$ for $k\in\bCp.$ A consequence of this
is the following. Since it is already known that
$f(k,x)$ and $f'(k,x)$ are analytic in $k\in\bCp,$
$f(-k,x)^\dagger$ and $f'(-k,x)^\dagger$ have analytic
extensions from $k\in\bR$ to $k\in\bCp$
given by $f(-k^*,x)^\dagger$ and $f'(-k^*,x)^\dagger,$ respectively.

Let $[F;G]:=FG'-F'G$ denote the Wronskian. It can directly be verified
that for any $n\times p$ solution $\psi(k,x)$ and
any $n\times q$ solution $\phi(k,x)$ to (1.1),
the Wronskians
$[\phi(k^*,x)^\dagger;\psi(k,x)]$ and
$[\phi(-k^*,x)^\dagger;\psi(k,x)]$ are both independent of $x.$
By evaluating the values of the Wronskians at $x=0$ and $x=+\infty,$
we can obtain various useful identities.
For example, we have
$$[f(\pm k,x)^\dagger;f(\pm k,x)]=
\pm 2ikI_n,\qquad k\in\bR,\tag 3.18$$
$$[f(-k^*,x)^\dagger;f(k,x)]=
0,\qquad k\in\bCpb.\tag 3.19$$

\vskip 10 pt
\noindent {\bf 4. THE JOST MATRIX AND THE SCATTERING MATRIX}
\vskip 3 pt

In this section we introduce the Jost matrix and the scattering
matrix for (1.1) with a selfadjoint matrix potential $V$ in
$L^1_1(\bR^+)$ and with the selfadjoint boundary condition
(1.11)-(1.13). We also present certain preliminary results
needed later on to analyze the small-$k$ limits of these two
matrices and of the inverse of the Jost matrix.

Recall that the Jost function $F_\theta$ corresponding to (1.3)
in the scalar case, i.e. when $n=1$ in (1.1), is defined with
the help of the Jost solution $f(k,x)$ as [5,16,29,31]
$$F_\theta(k):=\cases -i\left[f'(k,0)+(\cot\theta)\,f(k,0)\right],
\qquad \theta\in(0,\pi),\\
\stretch
f(k,0),\qquad \theta=\pi.\endcases\tag 4.1$$
We will define the matrix analog of the Jost function, which is
called the Jost matrix, so that it reduces
to the familiar Jost function when $n=1.$ Recall also that
the scattering matrix in the scalar case is defined as [5,16,29,31]
$$S_\theta(k):=\cases -\ds\frac{F_\theta(-k)}{F_\theta(k)},
\qquad \theta\in(0,\pi),\\
\stretch \ds\frac{F_\theta(-k)}{F_\theta(k)},\qquad
\theta=\pi.\endcases\tag 4.2$$ The reason behind the sign
difference in (4.2) in the Dirichlet case (i.e. when
$\theta=\pi$) is that (4.2) ensures that $S(k)\to 1$ as $V\to
0,$ which is a consequence of the fact that the perturbed and
unperturbed Hamiltonians satisfy the same selfadjoint boundary
condition at $x=0.$ We will define the scattering matrix by
generalizing (4.2) to the matrix case. For simplicity, we will
suppress the dependence of the Jost matrix and the scattering
matrix on the boundary-condition parametrization $(A,B),$ and
we will use the notation $J(k)$ for the Jost matrix instead of
$J_{(A,B)}(k)$ and also write $S(k)$ for the scattering matrix
instead of $S_{(A,B)}(k).$ Note that we earlier used $S(k,x)$
in (3.10) to denote the sine-like regular solution to (1.1),
which should not be confused with the notation $S(k)$ used for
the scattering matrix.

Define the Jost matrix $J(k)$ for $k\in\bCpb$ as
$$J(k):=[f(-k^*,x)^\dagger;\varphi(k,x)]=
f(-k^*,0)^\dagger B-f'(-k^*,0)^\dagger A,\tag 4.3$$ where
$f(k,x)$ is the Jost solution appearing in (3.1),
$\varphi(k,x)$ is the regular solution appearing in (3.6), and
$A$ and $B$ are the matrices appearing in (1.11)-(1.13) and
(3.6). Note that $J$ is not uniquely determined by the
potential $V$ and the selfadjoint boundary condition
(1.11)-(1.13). This is because (1.11)-(1.13) are invariant
under the transformation (1.17), and hence we have $J\mapsto
JD^\dagger$ under (1.17) indicating that the definition for $J$
in (4.3) is unique up to a right multiplication by a constant
invertible matrix. On the other hand, such a postmultiplication
does not change the zeros in $\bCp$ of the determinant of
$J(k).$ Those zeros correspond [1,19] to the bound-state
energies of (1.1) with the boundary condition (1.11)-(1.13),
and hence the bound-state energies are still uniquely
determined by (4.3).

\noindent {\bf Theorem 4.1} {\it If $V$ is selfadjoint
and belongs to $L^1_1(\bR^+),$ then the Jost matrix
$J(k)$ is invertible for $k\in\bR\setminus\{0\}$.}

\noindent PROOF: Even though a proof is available [19],
for the benefit of the reader we outline a proof of our own.
For $k\in\bR$ define
$$L(k):=f'(-k,0)^\dagger BE^{-2}+f(-k,0)^\dagger AE^{-2},\tag 4.4$$
where $E$ is the matrix $E_3$ appearing in (1.14).
With the help of (1.16), (3.18), (4.3), and (4.4) one can show that
$$J(k)\,L(k)^\dagger-L(k)\,J(k)^\dagger=
[f(-k,x)^\dagger;f(-k,x)]\big|_{x=0}=
-2ikI_n,\qquad k\in\bR.\tag 4.5$$
If $J(k)$ were noninvertible at some real nonzero $k_0,$ then the
rows of $J(k_0)$ would be linearly dependent and hence
we would have
$u^\dagger J(k_0)=0$ for some nonzero
vector $u\in\bold C^n$ as well as $J(k_0)^\dagger u=0.$ However, because
of (4.5) this would imply
$$0=u^\dagger J(k_0)\,L(k_0)^\dagger u-u^\dagger L(k_0)\,J(k_0)^\dagger u=-2ik_0
u^\dagger u=-2ik_0\,||u||^2\ne 0,$$
which is a contradiction. Thus, $J(k_0)$ must be invertible. \qed

The scattering matrix $S(k)$ is defined as [18-20]
$$S(k):=-J(-k)\,J(k)^{-1},\qquad k\in\bR\setminus\{0\},\tag 4.6$$
and it is uniquely determined by the boundary
condition and the potential $V.$
Even though $J(k)$ is uniquely defined only
up to a right multiplication by a constant invertible
matrix, the
unique determination of $S(k)$ is assured because
$S(k)$ remains invariant under the transformation (1.17).
Note that the domain of $J(k)$ is $k\in\bCpb$ because
$f(-k,0)^\dagger$ and $f'(-k,0)^\dagger$ have analytic
extensions from $k\in\bR$ to $k\in\bCp$ and the values
of those extensions are $f(-k^*,0)^\dagger$ and $f'(-k^*,0)^\dagger,$
respectively.
On the other hand, in general
$S(k)$ is defined only for real $k$ because $J(-k)$
in general cannot be extended from $k\in\bR$ to $k\in\bCp.$ Furthermore,
the existence of $S(k)$ when $k=0$
needs to be studied separately because, as we have seen in Theorem~4.1,
the existence of
$J(k)^{-1}$ is assured only for $k\in\bR\setminus\{0\}$
and it cannot easily be inferred from (4.6)
whether $S(k)$ has a limit
as $k\to 0$ when $J(0)^{-1}$ does not exist.

In order to understand the small-$k$ behavior of $J(k)$ and
$S(k),$ it is instructive to analyze first the case when the
potential $V$ is identically zero in (1.1). In that case, we
have $f(k,x)=e^{ikx}I_n,$ and hence (4.3) and (4.6) yield
$$J(k)=B-ikA, \quad [J(k)]^{-1}=(B-ikA)^{-1},
\quad S(k)=-(B+ikA)(B-ikA)^{-1}.$$ Let us use the
representation (2.6) for $(A,B)$ with the diagonal form of $U$
given in (2.7). We then obtain
$$A=-\text{diag}\{\sin\theta_1,\dots,\sin\theta_n\},
\quad B=\text{diag}\{\cos\theta_1,\dots,\cos\theta_n\},$$
$$J(k)=\text{diag}\{J_1(k),\dots,J_n(k)\},\quad S(k)=\text{diag}\{S_1(k),\dots,S_n(k)\},
$$
where we have defined
$$J_j(k):=\cos\theta_j+ik\,\sin \theta_j,\quad S_j(k):=\ds\frac{-\cos\theta_j+ik\,\sin \theta_j}{\cos\theta_j+ik\,\sin \theta_j}.\tag 4.7$$
As seen from (4.7), in the Dirichlet case
(i.e. when $\theta_j=\pi$) we have
$$J_j(k)=-1,\quad [J_j(k)]^{-1}=-1,\quad S_j(k)=-1.$$
On the other hand, in the Neumann case (i.e. when $\theta_j=\pi/2$) we have
$$J_j(k)=ik,\quad [J_j(k)]^{-1}=\ds\frac{1}{ik},\quad S_j(k)=1.$$
Note that, in the Neumann case, $J_j(0)$ vanishes linearly as $k\to 0$
and it is not an invertible matrix; however, $S_j(0)$ is still well defined because
$J(-k)[J(k)]^{-1}$ has a well-defined limit as $k\to 0.$
It is somehow disturbing that in the Dirichlet case,
$S_j(0)\neq 1$ and in fact $S_j(0)=-1,$ which is exactly the
opposite of the scalar case as seen from (4.2). The explanation
for the discrepancy is that the unperturbed Hamiltonian in the
matrix case is chosen to satisfy the Neumann boundary condition,
which
is compatible with the time-dependent derivation of the
scattering matrix and motivated by applications in
quantum wires; for further elaboration on this point we refer
the reader to p. 1566 of [24].

\vskip 10 pt \noindent {\bf 5. SMALL-$k$ BEHAVIOR} \vskip 3 pt

In preparation for the analysis of the small-energy
behavior of the Jost matrix $J(k),$ its inverse $J(k)^{-1},$ and
the scattering matrix $S(k),$
in this section we establish the small-$k$ asymptotics of
various quantities related to the regular solutions to (1.1).

We are interested in analyzing the Jost
matrix $J(k)$ as $k\to 0$ in $\bCpb.$ From (4.3) we see that
$$J(0)=f(0,0)^\dagger B-f'(0,0)^\dagger A,\tag 5.1$$
and we would like to determine how fast $J(k)$ approaches $J(0)$ and whether $J(k)^{-1}$ exists at $k=0$
and determine
its behavior as $k\to 0$ from $\bCpb.$ We would like to know about such
small-$k$ behaviors when $V$ is selfadjoint
and belongs to $L^1_1(\bR^+).$

As stated before (3.8), $f(k,a)$ is invertible
in the vicinity of $k=0$ in $\bCpb$ for some $a$ value.
In (4.3) we have defined the Jost matrix
in terms of a Wronskian whose value is independent of $x.$ As we see below we can write
$J(k)^\dagger$ in terms of Wronskians evaluated at $x=a$ and involving
the solutions $f(k,x),$ $\varphi(k,x),$ and $\omega(k,x)$
appearing in (3.1), (3.6), and (3.13), respectively.

The following result will be needed later on. By a generic constant, we mean a constant that does not necessarily
have the same value in different appearances.

\noindent {\bf Proposition 5.1} {\it If $V$ is selfadjoint
and belongs to $L^1_1(\bR^+),$ then the regular solution $\omega(k,x)$ to (1.1) appearing in (3.13) satisfies}
$$||\omega(k,x)-\omega(0,x)||\le c\left(\ds\frac{|k|(x-a)}
{1+|k|(x-a)}\right)^2 e^{(\text{Im}[k])(x-a)},
\qquad k\in\bCpb,\quad x\ge a,
\tag 5.2$$
{\it where $c$ is a generic constant.}

\noindent PROOF: From (3.11), (3.12), and (3.15)
we have
$$\omega(k,x)=f(0,a)\,\cos k(x-a)+f'(0,a)\,\ds\frac{\sin k(x-a)}{k}
+\ds\frac{1}{k}\int_a^x dy\,\sin k(x-y)\,V(y)\,\omega(k,y).\tag 5.3$$
Note that (5.3) yields
$$\omega(0,x)=f(0,a)+(x-a)\,f'(0,a)+\int_a^x dy\,(x-y)\,V(y)\,\omega(0,y),
\tag 5.4$$
and $\omega(0,x)=f(0,x)$ by (3.14). Thus, from (5.4) and its
$x$-derivative, with the help of $f'(0,x)=o(1/x)$ as $x\to+\infty,$ we obtain
$$\int_a^\infty dy\,V(y)\,\omega(0,y)=-f'(0,a),\tag 5.5$$
$$\int_a^\infty dy\,y\,V(y)\,\omega(0,y)=f(0,a)-a\,f'(0,a)-I_n.\tag 5.6$$
Let us write (5.5) as
$$f'(0,a)=-\int_a^x dy\,V(y)\,\omega(0,y)-\int_x^\infty dy\,V(y)\,\omega(0,y).\tag 5.7$$
Using (5.3), (5.4), and (5.7), we get
$$\omega(k,x)-\omega(0,x)=K_1+K_2+K_3+K_4,\tag 5.8$$
where we have defined
$$K_1:=f(0,a)\left[\cos k(x-a)-1\right],\tag 5.9$$
$$K_2:=\left[1-\ds\frac{\sin k(x-a)}{k(x-a)}\right] (x-a)\int_x^\infty dy\, V(y)\,\omega(0,y),\tag 5.10$$
$$K_3:=\ds\frac{1}{k}\int_a^x dy\,\left[\sin k(x-y)-k(x-y)-\sin k(x-a)+k(x-a)\right]\,V(y)\,\omega(0,y),\tag 5.11$$
$$K_4:=\ds\frac{1}{k}\int_a^x dy\,\left[\sin k(x-y)\right]\,V(y)\,\left[\omega(k,y)-\omega(0,y)\right].\tag 5.12$$
For $z\in\bCpb$ we have
$$|\sin z|\le \ds\frac{c\,|z|\,e^{\text{Im}[z]}}{1+|z|},\quad
\left|1-\ds\frac{\sin z}{z}\right|\le
\ds\frac{c\,|z|^2\,e^{\text{Im}[z]}}{(1+|z|)^2},\quad
\left|1-\cos z\right|\le
\ds\frac{c\,|z|^2\,e^{\text{Im}[z]}}{(1+|z|)^2} .\tag 5.13$$
Using (5.9) and the third estimate of (5.13), we get
$$||K_1||\le \ds\frac{c\,|k|^2(x-a)^2\,e^{(\text{Im}[k])(x-a)}}{(1+|k|(x-a))^2},
\qquad k\in\bCpb,\quad x\ge a.\tag 5.14
$$
Note that
$$\aligned \left\|(x-a)\int_x^\infty dy\,V(y)\,\omega(0,y)\right\|
\le &\left\|\int_x^\infty dy\,(y-a)\,V(y)\,\omega(0,y)\right\|
\\
\le &\left\|\int_x^\infty dy\,y\,V(y)\,\omega(0,y)\right\|.\endaligned\tag 5.15$$
The norms in (5.15) are bounded by a constant due to the facts that
$V\in L_1^1(\bR^+)$ and $\omega(0,x)$ is bounded as a result of (3.3) and (3.14).
Thus, from (5.10), (5.15), and the second estimate in (5.13), we get
$$||K_2||\le \ds\frac{c\,|k|^2(x-a)^2\,e^{(\text{Im}[k])(x-a)}}{(1+|k|(x-a))^2},
\qquad k\in\bCpb,\quad x\ge a.
\tag 5.16$$
Let us now estimate $K_3$
when $k\in\bCpb$ and $x\ge a.$
We write (5.11) as
$$K_3=\int_a^x dy\int_{x-y}^{x-a} dz\,\left[1-\cos kz\right]\,V(y)\,\omega(0,y),$$
and use the third estimate of (5.13) and the fact that $x\mapsto x^2/(1+x)^2$
is an increasing function of $x$ when $x\ge 0,$ to obtain
$$||K_3||\le \ds\frac{c\,|k|^2(x-a)^2\,e^{(\text{Im}[k])(x-a)}}{(1+|k|(x-a))^2}
\int_a^x dy\,y\,||V(y)||\,||\omega(0,y)||.\tag 5.17$$
Since  $\omega(0,y)$ is bounded and $V\in L^1_1(\bR^+),$ from
(5.17) we obtain
$$||K_3||\le \ds\frac{c\,|k|^2(x-a)^2\,e^{(\text{Im}[k])(x-a)}}{(1+|k|(x-a))^2},
\qquad k\in\bCpb,\quad x\ge a,\tag 5.18$$
for a generic constant $c.$
Let us now estimate $K_4$
when $k\in\bCpb$ and $x\ge a.$
Letting
$$\zeta(k,x):=e^{-(\text{Im}[k])(x-a)}||\omega(k,x)-\omega(0,x)||,\tag 5.19$$
 from (5.12) we get
$$e^{-(\text{Im}[k])(x-a)} ||K_4||\le
\ds\frac{1}{|k|}\int_a^x dy\,
e^{-(\text{Im}[k])(x-y)}
\left|\sin k(x-y)\right|\,||V(y)||\,\zeta(k,y)
.\tag 5.20$$
Using the first estimate of (5.13) in
(5.20) and the fact that $x\mapsto x/(1+x)$
is an increasing function of $x$ when $x\ge 0,$
we obtain
$$e^{-(\text{Im}[k])(x-a)}
||K_4||\le \ds\frac{c(x-a)}{(1+|k|(x-a))} \int_a^x
dy\,||V(y)||\,\zeta(k,y).\tag 5.21$$ Using (5.14), (5.16),
(5.18), (5.19), and (5.21) in (5.8) we get
$$\zeta(k,x)\le \ds\frac{c\,|k|^2(x-a)^2}{(1+|k|(x-a))^2}+
\ds\frac{c}{|k|}\ds\frac{|k|(x-a)}{(1+|k|(x-a))}
\int_a^x dy\,||V(y)||\,\zeta(k,y).\tag 5.22$$
Setting
$$\chi(k,x):=\ds\frac{(1+|k|(x-a))^2}{c\,|k|^2(x-a)^2}\,\zeta(k,x),
\tag 5.23$$
we can write (5.22) as
$$\chi(k,x)\le 1+\ds\frac{(1+|k|(x-a))}{|k|^2(x-a)}
\int_a^x dy\,||V(y)||\,\ds\frac{c\,|k|^2(y-a)^2}{(1+|k|(y-a))^2}
\,\chi(k,y).\tag 5.24$$
 From (5.24) we obtain
$$\chi(k,x)\le 1+c\int_a^x dy\,y \,||V(y)||\,\chi(k,y),\tag 5.25$$
where we have used, for $0\le a\le y\le x,$ the estimate
$$\ds\frac{|k|^2(y-a)^2}{(1+|k|(y-a))^2}
\le
\ds\frac{|k|(x-a)}{(1+|k|(x-a))}
\ds\frac{|k|(y-a)}{(1+|k|(y-a))}
\le
\ds\frac{|k|(x-a)}{(1+|k|(x-a))}\,|k|\,y,
$$
based on the fact that
$x\mapsto x/(1+x)$
is an increasing function of $x$ when $x\ge 0.$
Applying Gronwall's lemma to (5.25) and using the fact that
$V\in L^1_1(\bR^+),$ we obtain
$\chi(k,x)\le c$ for some generic constant $c,$ which is not
necessarily equal to the generic constant $c$ in (5.25).
Thus, using (5.19) and (5.23) in $\chi(k,x)\le c$ we obtain
(5.2). \qed

Let us define
$$P(k):=[\omega(-k^*,x)^\dagger;f(k,x)],\tag 5.26$$
where we note that
the Wronskian in (5.26) is independent of $x,$ and hence
with the help of (3.13) and (3.16) by evaluating that Wronskian at $x=a$ we get
$$P(k)=f(0,a)^\dagger f'(k,a)-f'(0,a)^\dagger f(k,a).\tag 5.27$$
Note that $P(k)$ has an analytic extension
 from $k\in\bR$ to $k\in\bCp$
because $f(k,a)$ and $f'(k,a)$ possess that property as well.
It is difficult to obtain useful information from (5.27)
as $k\to 0$ because for $V\in L^1_1(\bR^+)$ we can only say that
$$f(k,x)=f(0,x)+o(1),\quad f'(k,x)=f'(0,x)+o(1),\qquad k\to 0 \text{ in }
\bCpb.\tag 5.28$$
In the proposition below we evaluate the small-$k$ asympotics of $P(k)$
by evaluating the Wronskian in (5.26) at
$x=+\infty.$ This result will be useful in evaluating
the small-$k$ limit of the Jost matrix $J(k).$

\noindent {\bf Proposition 5.2} {\it If $V$ is selfadjoint
and belongs to $L^1_1(\bR^+),$ then the matrix $P(k)$ given in (5.26) satisfies}
$$P(k)=ikI_n+o(k),\qquad k\to 0 \text{ in }\bCpb.\tag 5.29$$

\noindent PROOF:
We can evaluate the asymptotics of
$\omega(-k^*,x)$ and $\omega'(-k^*,x)$ as $x\to+\infty$ from (5.3). Furthermore,
the asymptotics of $f(k,x)$ and $f'(k,x)$
as $x\to+\infty$ are available from (3.1). Using those
asymptotics in (5.26) we obtain
$$P(k)=ike^{ika}\,f(0,a)^\dagger-e^{ika}\,f'(0,a)^\dagger
-\int_a^\infty dy\,e^{iky}\,\omega(-k^*,y)^\dagger\,V(y).\tag 5.30$$
Let
us break the right hand side in (5.30) into three terms and
write
$$P(k)=P_1(k)+P_2(k)+P_3(k),\tag 5.31$$
where we have defined
$$P_1(k):=ike^{ika}\,f(0,a)^\dagger-e^{ika}\,f'(0,a)^\dagger,\tag 5.32$$
$$P_2(k):=-\int_a^\infty dy\,e^{iky}\,[\omega(-k^*,y)^\dagger-\omega(0,y)^\dagger]
\,V(y),\tag 5.33$$
$$P_3(k):=-\int_a^\infty dy\,e^{iky}\,\omega(0,y)^\dagger\,V(y).\tag 5.34$$
 From (5.2) and the Lebesgue dominated convergence
theorem, it follows
that $P_2(k)=o(k)$ as $k\to 0$ in $\bCpb$ whenever
$V\in L_1^1(\bR^+).$ Using (5.5) and (5.6) in (5.34)
we have
$$P_3(k)=f'(0,a)^\dagger+ik[I_n+a\,f'(0,a)^\dagger-f(0,a)^\dagger]+P_4(k),
\tag 5.35$$
where we have defined
$$P_4(k):=-\int_a^\infty dy\,[e^{iky}-1-iky]\,\omega(0,y)^\dagger
\,V(y).\tag 5.36$$
Note that for any $z\in\bCpb$ we have
$$|e^{iz}-1|\le c\,|z|,\quad |e^{iz}-1-iz|\le \ds\frac{c\,|z|^2}{1+|z|},
\tag 5.37$$
where $c$ denotes a generic nonnegative
constant independent of the complex number $z,$
and hence the second inequality in (5.37)
helps us to get $P_4(k)=o(k)$ as $k\to 0$ in $\bCpb.$
Thus, using (5.31)-(5.36), we have
$$P(k)=ikI_n+[ik e^{ika}-ik]\,f(0,a)^\dagger+
[I_n+ika- e^{ika}]\,f'(0,a)^\dagger+o(k),$$
and because of (5.37) each of the coefficients of $f(0,a)^\dagger$
and $f'(0,a)^\dagger$ is $O(k^2)$ as $k\to 0$ in $\bCpb.$ \qed

Even though (5.28) does not provide much extraordinary
information, the following theorem
shows that $f'(k,x)\,[f(k,x)]^{-1}$ is differentiable
at $k=0$ at any fixed $x$ value where the matrix $f(0,x)$ is invertible.
The results stated in the next theorem are the generalization
to the matrix case of similar results in the scalar case [2,3].

\noindent {\bf Theorem 5.3} {\it Assume that $V$ in (1.1)
is selfadjoint
and belongs to $L^1_1(\bR^+).$ If the constant matrix
$f(0,a)$ is invertible, where $f(k,x)$ is the Jost solution appearing
in (3.1), then, $f'(k,a)\,[f(k,a)]^{-1}$ is differentiable
at $k=0$ and we have}
$$f'(k,a)\,f(k,a)^{-1}=f'(0,a)\,f(0,a)^{-1}+ik[
f(0,a)^{-1}]^\dagger f(0,a)^{-1}+o(k),\qquad k\to 0 \text{ in
}\bCpb.\tag 5.38$$ {\it If instead, $f'(0,a)$ is invertible at
some $a\in\bR^+,$ then the Jost solution satisfies}
$$f(k,a)\,f'(k,a)^{-1}=f(0,a)\,f'(0,a)^{-1}-ik[
f'(0,a)^{-1}]^\dagger f'(0,a)^{-1}+o(k),\qquad k\to 0 \text{ in }\bCpb.\tag 5.39$$

\noindent PROOF: In case $f(0,a)$ is invertible, by the continuity
of the determinant of $f(k,a),$ we must have $f(k,a)$ invertible
in $\bCpb$ in the vicinity of $k=0.$ Thus, from (5.27) we get
$$f'(k,a)\,f(k,a)^{-1}=[f(0,a)^\dagger]^{-1}f'(0,a)^\dagger+
[f(0,a)^\dagger]^{-1}P(k)\,f(k,a)^{-1}.\tag 5.40$$
Note that (3.19) implies that
$$[f(0,a)^\dagger]^{-1}f'(0,a)^\dagger=f'(0,a)\,f(0,a)^{-1}.\tag 5.41$$
Applying (5.29) in (5.40) and using (3.8) and (5.41),
we obtain (5.38).
In a similar way, (5.39) is proved. \qed

Next, we express the Jost matrix
$J(k)$ defined in (4.3) in terms of the Jost solution $f(k,x),$
the regular solutions $\varphi(k,x),$ the regular solution
$\omega(k,x),$ and the matrix $P(k)$
appearing in (3.1), (3.6), (3.13), and (5.26), respectively.

\noindent {\bf Proposition 5.4} {\it Assume that $V$ in (1.1)
is selfadjoint
and belongs to $L^1_1(\bR^+).$ The Jost matrix can be written as}
$$J(k)=T_1(k)+T_2(k),\qquad k\in\bCpb,\tag 5.42$$
{\it where we have defined}
$$T_1(k):=-P(-k^*)^\dagger f(0,a)^{-1}\,\varphi(k,a),\tag 5.43$$
$$T_2(k):=f(-k^*,a)^\dagger [f(0,a)^{-1}]^\dagger
[\omega(-k^*,x)^\dagger;\varphi(k,x)],
\tag 5.44$$
{\it and recall that the value
of the Wronskian appearing in (5.44) is independent of $x.$}

\noindent PROOF: Using (4.3) we can write $J(k)$ in terms of
Wronskians evaluated at $x=a$ to get
$$\aligned
J(k)=&[f(-k^*,x)^\dagger;\varphi(k,x)]\big|_{x=a}\\
=&f(-k^*,a)^\dagger \varphi'(k,a)- f'(-k^*,a)^\dagger \varphi(k,a)  \\
=&T_1(k)+T_2(k)+T_3(k),\endaligned$$
where we have defined
$$T_1(k):=- \left(
f'(-k^*,a)^\dagger f(0,a)-f(-k^*,a)^\dagger f'(0,a)
\right)f(0,a)^{-1}\varphi(k,a),\tag 5.45$$
$$T_2(k):=f(-k^*,a)^\dagger [f(0,a)^{-1}]^\dagger\left(
f(0,a)^\dagger \varphi'(k,a)-f'(0,a)^\dagger \varphi(k,a)
\right),\tag 5.46$$
$$T_3(k):=f(-k^*,a)^\dagger [f(0,a)^{-1}]^\dagger
\left(f'(0,a)^\dagger f(0,a)-f(0,a)^\dagger
f'(0,a)\right)
f(0,a)^{-1} \varphi(k,a)
.\tag 5.47$$
Using (5.27) in (5.45) we see that $T_1(k)$ in (5.45)
can equivalently be written as in (5.43). Using (3.13)
in (5.46), we see that $T_2(k)$ in (5.46)
can equivalently be written as
$$\aligned
T_2(k)=&f(-k^*,a)^\dagger [f(0,a)^{-1}]^\dagger\left(
\omega(-k^*,a)^\dagger \varphi'(k,a)-\omega'(-k^*,a)^\dagger \varphi(k,a)
\right)\\
=&f(-k^*,a)^\dagger [f(0,a)^{-1}]^\dagger
 [\omega(-k^*,x)^\dagger;
\varphi(k,x)]\big|_{x=a}
,\endaligned$$
which is equivalent to (5.44) because the Wronskian
in (5.44) is independent of $x$ and can be evaluated at $x=a.$
Finally, using (3.19) with $k=0$ in (5.47)
we see that $T_3(k)=0.$ \qed

In the next theorem we evaluate the Wronskian $[\omega(-k^*,x)^\dagger;\varphi(k,x)]$
appearing in (5.44).

\noindent {\bf Proposition 5.5} {\it Assume that $V$ in (1.1)
is selfadjoint
and belongs to $L^1_1(\bR^+).$ Then, the Wronskian appearing in
(5.44) has the small-$k$ asymptotics}
$$[\omega(-k^*,x)^\dagger;\varphi(k,x)]=J(0)+O(k^2),
\qquad k\to 0 \text{ in } \bCpb,\tag 5.48$$
{\it where $J(k)$ is the Jost matrix defined in (4.3).}

\noindent PROOF: Since the value of the
Wronskian in (5.48) is independent of
$x,$ we will evaluate its value at $x=0.$
By writing
$$[\omega(-k^*,x)^\dagger;\varphi(k,x)]=
[\omega(-k^*,x)^\dagger-\omega(0,x)^\dagger;\varphi(k,x)]+
[\omega(0,x)^\dagger;\varphi(k,x)],\tag 5.49$$
 from (3.6), (3.14), and (5.1)
we see that the second Wronskian on the right side in (5.49), when
$x=0,$  yields
$$[\omega(0,x)^\dagger;\varphi(k,x)]\big|_{x=0}=J(0).$$
Next, we evaluate at $x=0$ the value of the
first Wronskian on the right side in (5.49).
The $x$-derivative of
that first Wronskian,
with the help of (1.1) and (3.17), can be directly evaluated
as
$$\ds\frac{d}{dx}[\omega(-k^*,x)^\dagger-\omega(0,x)^\dagger; \varphi(k,x)]
=k^2\,\omega(0,x)^\dagger\,\varphi(k,x).\tag 5.50$$
Integrating (5.50) over the interval
$[0,a],$ and then using (3.13) and (3.14), we obtain
$$[\omega(-k^*,x)^\dagger-\omega(0,x)^\dagger;\varphi(k,x)]
\big|_{x=0}=-k^2 \int_0^a
dy\,f(0,y)^\dagger\,\varphi(k,y) .$$
Thus, we have the estimate in (5.48). \qed

Using (5.4), (5.29), (5.42)-(5.44), and (5.48) we have the following
conclusion.

\noindent {\bf Corollary 5.6} {\it Assume that $V$ in (1.1)
is selfadjoint
and belongs to $L^1_1(\bR^+).$ Then, the Jost matrix $J(k)$ appearing in
(4.3) has the small-$k$ behavior}
$$f(0,a)^\dagger [f(-k^*,a)^\dagger]^{-1}
J(k)=J(0)-ik\,f(0,a)^{-1}\,
\varphi(0,a)+o(k),
\qquad k\to 0 \text{ in } \bCpb,\tag 5.51$$
{\it where $f(k,x)$ and
$\varphi(k,x)$ are the Jost solution and the regular
solution appearing
in (3.1) and (3.6), respectively,
and $a$ is any point where the matrix
$f(0,a)$ is invertible.}

In order to study the small-$k$ limit of
$J(k)^{-1},$ we will next concentrate on the $O(k)$-term
appearing in (5.51), namely $f(0,a)^{-1}\,\varphi(0,a).$
We use $\text{Ker}\, [J(0)]$ to denote
the kernel of the matrix $J(0).$

\noindent {\bf Proposition 5.6} {\it Assume that $V$ in (1.1)
is selfadjoint
and belongs to $L^1_1(\bR^+).$ Then, the following are equivalent:}

\item {(a)} {\it  The vector
$u\in \bold C^n$ is an eigenvector of the zero-energy Jost matrix $J(0)$
with the zero eigenvalue, i.e.
$u\in \text{Ker}\, [J(0)].$}

\item {(b)} {\it $\varphi'(0,+\infty)\,u=0,$ where $\varphi(k,x)$
is the regular solution to (1.1) appearing in (3.6).}

\item {(c)} {\it $\varphi(0,x)\,u$ is bounded for $x\in[0,+\infty).$}

\noindent PROOF: From (4.3) we see that
$$J(0)=f(0,x)^\dagger\,\varphi'(0,x)-f'(0,x)^\dagger\,\varphi(0,x),\tag 5.52$$
where the quantity on the right side in (5.52) is independent of $x.$
 From (3.5) it follows that each column of
$\varphi(0,x)$ is a linear combination of
columns of $f(0,x)$ and $g(0,x).$ Hence, there exist constant
$n\times n$ matrices
$\alpha$ and $\beta$ such that
$$\varphi(0,x)=f(0,x)\,\alpha+g(0,x)\,\beta,\qquad x\in\bR^+.\tag 5.53$$
 From the $x$-derivative of (5.53) we get
$$\varphi'(0,x)=f'(0,x)\,\alpha+g'(0,x)\,\beta,\qquad x\in\bR^+.\tag 5.54$$
Using (3.3) and (3.4) in (5.54) we see that
$$\varphi'(0,+\infty)=\beta.\tag 5.55$$
Inserting (5.53) and (5.54) on the right side of (5.52),
evaluating the resulting expression as $x\to+\infty,$ and using
(3.3) and (3.4), we obtain
$$J(0)=\beta.\tag 5.56$$
Thus, from (5.55) and (5.56) we conclude that
$$J(0)=\varphi'(0,+\infty),\tag 5.57$$
and hence the equivalence of (a) and (b) are established.
Note that, from (3.3), (3.4), and (5.53) it follows that
$\varphi(0,x)\,u$ is bounded if and only if $\beta u=0,$
which happens if and only if $\varphi'(0,+\infty)\,u=0$
as a result of (5.55). Thus, the equivalence of (b) and (c)
is established. \qed

Let us note that we can express $\varphi'(0,+\infty)$
in (5.57) in another form.
Letting $k\to 0$ in (3.7) we get
$$\varphi(0,x)=A+Bx+\int_0^x dy\,(x-y)\,V(y)\,\varphi(0,y),\tag 5.58$$
and from the $x$-derivative of (5.58) we have
$$\varphi'(0,x)=B+\int_0^x dy\,V(y)\,\varphi(0,y),\tag 5.59$$
We know from (3.3)-(3.5) that $\varphi(0,x)$ can grow at most
as $O(x)$ as $x\to +\infty$ and hence the integral in (5.59)
exists as $x\to+\infty,$ and from (5.57) and (5.59) we get
$$J(0)=\varphi'(0,+\infty)=B+\int_0^\infty dy\,V(y)\,\varphi(0,y).$$

\noindent {\bf Proposition 5.7} {\it Assume that $V$ in (1.1)
is selfadjoint
and belongs to $L^1_1(\bR^+).$ Then, for any vector
$u$ in $\text{Ker}\, [J(0)]$ there
exists a unique vector $\xi$ in $\text{Ker}\, [J(0)^\dagger]$
such that}
$$\varphi(0,x)\,u=f(0,x)\,\xi.\tag 5.60$$
{\it The map $u\mapsto\xi$ from
$\text{Ker}\, [J(0)]$ to $\text{Ker}\, [J(0)^\dagger]$ is a bijection.}

\noindent PROOF: By expressing $\varphi(0,x)\,u$ as in (3.5) we get
$$\varphi(0,x)\,u=f(0,x)\,\xi+g(0,x)\,\eta,$$
and hence $\varphi(0,x)\,u$ is bounded,
equivalently as stated in Proposition~5.6, $u\in \text{Ker}\, [J(0)],$
if and only if
$\eta=0.$ Thus, the mapping $u\mapsto\xi$ is identified with (5.44).
Let us now show that $\xi\in \text{Ker}\, [J(0)^\dagger].$ Using (5.1),
we have
$$J(0)^\dagger\xi=[B^\dagger f(0,0)-A^\dagger f'(0,0)]\xi.\tag 5.61$$
On the other hand, from (5.44) and its derivative, with the
help of (3.6), we see that
$$f(0,0)\,\xi=\varphi(0,0)\,u=Au,\quad f'(0,0)\,\xi=\varphi'(0,0)\,u=Bu.\tag 5.62$$
Using (5.62) in (5.61) and imposing (1.12) we get $J(0)^\dagger\xi=0.$
Thus, $\xi\in \text{Ker}\, [J(0)^\dagger].$ Note that the map
$u\mapsto \xi$ is a linear map from $\text{Ker}\, [J(0)]$ into
$\text{Ker}\, [J(0)^\dagger]$ because, as seen from (5.60), we have
$$\xi=f(0,a)^{-1}\varphi(0,a)\,u,
\qquad u\in \text{Ker}\, [J(0)].\tag 5.63$$
The map $u\mapsto\xi$ from
$\text{Ker}\, [J(0)]$ to $\text{Ker}\, [J(0)^\dagger]$
is one-to-one because it has zero kernel as seen by the
following argument. If $\xi=0$ in (5.60), then we must have
$\varphi(0,x)\,u=0$ and $\varphi'(0,x)\,u=0.$ In particular, at $x=0$ with the help
of (3.6) we then have $Au=0$ and $Bu=0.$ We in turn get
$A^\dagger Au=0$ and $B^\dagger Bu=0,$ and hence
$(A^\dagger A+B^\dagger B)u=0,$
which yields $u=0$ because of (1.13).
Furthermore, $\text{Ker}\, [J(0)]$ and $\text{Ker}\, [J(0)^\dagger]$
have the same dimension, and hence the map $u\mapsto\xi$ from
$\text{Ker}\, [J(0)]$ to $\text{Ker}\, [J(0)^\dagger]$
is a bijection. \qed

\vskip 10 pt \noindent {\bf 6. SMALL-$k$ BEHAVIOR OF
$J(k)^{-1}$ AND OF $S(k)$} \vskip 3 pt

In this section we establish
the small-$k$ asymptotics of the Jost matrix
$J(k),$ its inverse $J(k)^{-1},$ and the scattering matrix
$S(k).$ As we will see, $J(k)$ is continuous at $k=0,$
$J(k)^{-1}$ has an $O(1/k)$ singularity at $k=0$ if
$J(0)$ has a zero eigenvalue,
$J(k)$ is continuous at $k=0$ if zero is not an
eigenvalue of $J(0),$ and that
$S(k)$ is continuous at $k=0$ whether or not zero is an eigenvalue
of $J(0).$

In order to analyze the small-$k$ behavior of
$J(k)^{-1},$ we will analyze (5.51). Let us write
(5.51) as
$$F(k)=J(0)-ik R+o(k),\qquad k\to 0\text{ in }
\bCpb,\tag 6.1$$
where we have defined
$$F(k):=f(0,a)^\dagger [f(-k^*,a)^{-1}]^\dagger
J(k),\quad R:=f(0,a)^{-1}\,
\varphi(0,a).\tag 6.2$$
We will equivalently analyze the behavior of
$F(k)^{-1}$ as $k\to 0$ $\in\bCpb.$
As we have seen in Proposition~5.7, the restriction
of $R$ to $\text{Ker}\, [J(0)]$ yields an invertible
map. Among the $n$ eigenvalues of
$J(0),$ let us assume that the zero eigenvalue has
geometric multiplicity $\mu$ with a possibly larger
algebraic multiplicity $\nu.$ In other words,
$J(0)$ has $\mu$ linearly independent
eigenvectors corresponding to the
zero eigenvalue. Thus, $J(0)\,v=0$
if and only if $v\in\bold C^n$ is such an eigenvector,
in which case we have $Rv\ne 0.$

Let us choose a Jordan basis [10] for the matrix $J(0)$ as follows.
Assume that there are $\kappa$ Jordan chains and hence
$\kappa$ blocks in the Jordan canonical form of $J(0).$ Let
us use the index $\alpha$ for $\alpha=1,\dots,\kappa$
to identify the Jordan chains and assume that
the $\alpha$th chain consists
of $n_\alpha$ vectors $u_{\alpha j}$ for $j=1,\dots,n_\alpha.$
Let us use $\lambda_\alpha$ to denote the eigenvalue of
$J(0)$ associated with the $\alpha$th chain, where the eigenvalues
may be repeated. We have
$$\cases [J(0)-\lambda_\alpha]\,u_{\alpha 1}=0,\\
[J(0)-\lambda_\alpha]\,u_{\alpha j}=u_{\alpha (j-1)},\qquad
j=2,\dots,n_\alpha,\endcases\tag 6.3$$
and hence $u_{\alpha 1}$ is an eigenvector and
$u_{\alpha j}$ for $j=2,\dots,n_\alpha$ are the generalized
eigenvectors.

Since we assume the zero eigenvalue
has
geometric multiplicity $\mu,$ without loss of generality
we let $\lambda_\alpha=0$ for $\alpha=1,\dots,\mu$ and
$\lambda_\alpha\ne 0$ for $\alpha=\mu+1,\dots,\kappa.$
We order the vectors in the Jordan basis
according to the rule that $u_{\alpha j}$ comes before
$u_{\beta s}$ if and only if $\alpha<\beta$ or $\alpha=\beta$
and $j<s.$ Thus, $\{u_{11},u_{21},\dots,u_{\mu 1}\}$
forms a basis for $\text{Ker}\, [J(0)],$
and our Jordan basis is given by the ordered set $\{u_{\alpha j}\},$ i.e.
$$\{u_{11},u_{12},\dots,u_{1 n_1},u_{21},u_{22},\dots,u_{2
n_2},\dots,
u_{\kappa 1},u_{\kappa 2},\dots,u_{\kappa n_\kappa}\}.\tag 6.4$$

The corresponding adjoint Jordan basis $\{v_{\alpha j}\}$
satisfies $v_{\alpha j}^\dagger u_{\rho t}=\delta_{\alpha
\rho}\delta_{jt},$ with $\delta_{jt}$ denoting the Kronecker
delta, and the indices $\alpha$ and $\rho$ referring to the
Jordan blocks. The vectors $v_{\alpha j}$ satisfy
$$\cases [J(0)^\dagger-\lambda^*_\alpha]\,v_{\alpha n_\alpha}=0,\\
[J(0)^\dagger-\lambda^*_\alpha]\,v_{\alpha j}=u_{\alpha (j+1)},\qquad
j=1,\dots,n_\alpha-1.\endcases\tag 6.5$$
Thus, $\{v_{1 n_1},v_{2 n_2},\dots,v_{\mu n_\mu}\}$
forms a basis for $\text{Ker}\, [J(0)^\dagger].$
The adjoint Jordan basis is the
ordered set $\{v_{\alpha j}\},$ i.e.
$$\{v_{11},v_{12},\dots,v_{1 n_1},v_{21},v_{22},\dots,v_{2
n_2},\dots,
v_{\kappa 1},v_{\kappa 2},\dots,v_{\kappa n_\kappa}\}.\tag 6.6$$

Let $\Cal S$ denote the matrix whose columns are given by the
elements of the ordered set $\{u_{\alpha j}\}$ in (6.4).
Then, $\Cal S^{-1}$ is exactly the matrix whose rows are given by the
elements of the ordered set $\{v^\dagger_{\alpha j}\},$
with the ordering given in (6.6).
Thus, the Jordan canonical form of $J(0)$ is given by
$$\Cal S^{-1}J(0)\,\Cal S=\oplus_{\alpha=1}^\kappa J_{n_\alpha}(\lambda_\alpha),\tag 6.7$$
where $J_{n_\alpha}(\lambda_\alpha)$ is the $n_\alpha\times
n_\alpha$ Jordan block
with $\lambda_\alpha$ appearing in the diagonal entries
and one in the superdiagonal entries.
Since the first $\mu$ Jordan blocks $J_{n_\alpha}(\lambda_\alpha)$
are associated with the zero eigenvalue
and the remaining $(n-\mu)$ blocks are associated
with nonzero eigenvalues, each $J_{n_\alpha}(\lambda_\alpha)$
is an $n_\alpha\times n_\alpha$ matrix given by
$$J_{n_\alpha}(\lambda_\alpha)=\bm
0&1&0&\dots &0&0\\
\stretch
0&0&1&\dots&0&0\\
\stretch
\vdots&\vdots&\vdots&\ddots&\vdots&\vdots\\
0&0&0&\dots&0&1\\
\stretch
0&0&0&\dots&0&0\endbm,\qquad \alpha=1,\dots,\mu,\tag 6.8$$
$$J_{n_\alpha}(\lambda_\alpha)=\bm
\lambda_\alpha&1&0&\dots &0&0\\
\stretch
0&\lambda_\alpha&1&\dots&0&0\\
\stretch
\vdots&\vdots&\vdots&\ddots&\vdots&\vdots\\
0&0&0&\dots&\lambda_\alpha&1\\
\stretch
0&0&0&\dots&0&\lambda_\alpha\endbm,\qquad \alpha=\mu+1,\dots,\kappa.\tag 6.9$$

Let us use a tilde
to denote the transformation via $\Cal S,$ i.e.
$\tilde M:=\Cal S^{-1} M\Cal S$ for any $n\times n$ matrix $M.$
Let us apply this transformation on the matrix
$F(k)$ appearing in (6.1) and (6.2). Then (6.1) yields
$$\tilde F(k)=\tilde J(0)-ik\tilde R+o(k),\qquad k\to 0\text{ in }
\bCpb.\tag 6.10$$
By inspecting (6.7)-(6.10) we see that
there are exactly $\mu$ columns of $\tilde F(k)$
behaving as $O(k)$ as $k\to 0$ and each of the remaining $(n-\mu)$
column vectors contains at least one entry that has a nonzero
limit as $k\to 0.$

Our next goal is to move all the entries with $1$ appearing in the superdiagonal
in the first $\mu$ Jordan blocks in (6.8) and collect all those entries into
the $(\nu-\mu)$ identity matrix $I_{\nu-\mu}.$ Recall that
$\nu$ and $\mu$ correspond to
the algebraic and geometric multiplicities of the zero
eigenvalue of $J(0),$ and hence there are exactly $(\nu-\mu)$ such entries to move.
Such a movement will be accomplished by first permuting
some of the first $\nu$ columns in $\tilde J(0)$ and then by permuting some of the
first $\nu$ rows of the resulting matrix.
The permutations among the first $\nu$ columns can be described by a matrix
postmultiplying $\tilde J(0)$ and we use $P_1$ to denote that matrix.
On the other hand, the permutations among the first $\nu$ rows can be described by a matrix
premultiplying $\tilde J(0)$ and we use $P_2$ to denote that matrix.
Thus, the matrix $P_2\tilde J(0)\,P_1$ will be given by
$$P_2\tilde J(0)\,P_1=\text{diag}\{0_\mu, I_{\nu-\mu}, J_{n_\mu+1}(\lambda_{\mu+1})
,\dots
,J_{n_\kappa}(\lambda_{\kappa})\},$$
where $0_\mu$ denotes the $\mu\times\mu$ zero matrix.
Since $P_1$ and $P_2$ affect only the first $\nu$ columns and $\nu$ rows, respectively,
they have the form
$$P_1=\bm \Pi_1&0\\
\stretch
0&I_{n-\nu}\endbm,\quad
P_2=\bm \Pi_2&0\\
\stretch
0&I_{n-\nu}\endbm,\tag 6.11$$
for some permutation matrices $\Pi_1$ and $\Pi_2.$

Formally speaking, the matrix $\Pi_1$ describes the permutation
$\pi_1$ given by
$$\pi_{1}: (1,\dots,\nu) \mapsto (q_{1},\dots,q_{\nu}),$$
where
$$q_{\tau}=\cases n_{1}+\cdots +n_{\tau-1}+1,
\quad &\tau =1,\dots,\mu, \\
\noalign{\medskip}
\tau-\mu+\alpha,
&\tau =\mu+1,\dots,\nu,\endcases$$
and $\alpha \in \{1,\dots, \mu \}$ is the unique integer
such that, for given $\tau$ and $\mu,$
$$n_{1}+n_{2}+\cdots +n_{\alpha-1}-\alpha+j=\tau -\mu,$$
for some $j \in
\{2,\dots ,n_{\alpha} \}.$ Note that, since $n_{\alpha} \ge 1,$
the quantity $n_{1}+n_{2}+\cdots
+n_{\alpha-1}-\alpha$ is a nondecreasing function of $\alpha.$

Similarly, $\Pi_2$ is related to the permutation
$\pi_{2}$ given by
$$\pi_{2}: (1,\dots, \nu) \mapsto
(\sigma_{1} ,\dots,\sigma_{\nu}),$$
where
$$\sigma_{\alpha}=\cases n_{1}+\cdots +n_{\alpha}, \quad
&\alpha =1,\dots,\mu, \\
\noalign{\medskip}
\alpha-\mu+\rho-1, &\alpha =\mu+1,\dots,\nu, \endcases$$
and $\rho \in \{1,\dots, \mu \}$ is the unique integer such that,
for given $\alpha$ and $\mu$
$$n_{1}+n_{2}+\cdots +n_{\rho-1}-\rho+s=\alpha -\mu,$$
for some $s \in
\{2,\dots ,n_{\rho} \}.$
To implement these permutations we let $\hat e_{j}$ for $j=1,\dots, \nu$
denote the column vectors of the standard basis in $\bold C^{\nu}$
and let $\Pi_{1}$ be the $\nu \times \nu$ permutation matrix whose $j$th
column vector is $\hat e_{q_{j}},$ and let $\Pi_{2}$ be the $\nu
\times \nu$ permutation matrix whose $k$th row vector is $\hat
e_{\sigma_{k}}^{\dagger}.$  Now observe that, if $M$ is any $\nu \times
\nu$ matrix, then the matrix $\Pi_{2}\, M\, \Pi_{1}$ can be thought of
as being obtained from $M$ by a permutation of the columns according
to $\pi_{1}$ and a permutation of the rows according to $\pi_{2}.$

Let us now return to the matrix $F(k)$ defined in (6.2). By
first putting it into the Jordan canonical form $\tilde F(k)$ and
then by applying $P_1$ and $P_2$ on the first $\nu$ columns and rows
of $\tilde F(k),$
we form the matrix $\Cal Z(k)$ defined as
$$\Cal Z(k):=
\bm \Cal A(k)&\Cal B(k)\\
\stretch
\Cal C(k)&\Cal D(k)\endbm:
=P_2\, \tilde F(k)\,P_1=P_2\,\Cal S^{-1}F(k)\,\Cal S\,P_1,
\tag 6.12$$
where $\Cal A(k)$ has size $\mu\times \mu,$
$\Cal D(k)$ has size $(n-\mu)\times (n-\mu),$
$\Cal A(k)$ coincides with the submatrix
of $\tilde F(k)$ consisting of
the entries in columns $\alpha 1$ and rows $s n_s,$
where $\alpha=1,\dots,\mu$ and $s=1,\dots,\mu.$
The procedure of going from $F(k)$ to
$\Cal Z(k)$ is similar to
the procedure described on pp. 4638--4639
of [4], where the mappings
$P_1$ and $P_2$ were also used.

The small-$k$ limits of the block entries
in the matrix $\Cal Z(k)$ are described in the following theorem.

\noindent {\bf Theorem 6.1} {\it Assume that $V$ in (1.1)
is selfadjoint
and belongs to $L^1_1(\bR^+).$ Then, the asymptotics
as $k\to 0$ in $\bCpb$ of the
matrices $\Cal A(k),$ $\Cal B(k),$ $\Cal C(k),$ $\Cal D(k)$
appearing in (6.12) are given by}
$$\Cal A(k)=k \Cal A_1+o(k),
\quad \Cal B(k)=k \Cal B_1+o(k),
\quad \Cal C(k)=k \Cal C_1+o(k),
\quad \Cal D(k)=\Cal D_0+O(k),
\tag 6.13$$
{\it where $\Cal A_1,$ $\Cal B_1, $ $\Cal C_1,$ $\Cal D_0$
are constant matrices, and furthermore
$\Cal A_1$ and $\Cal D_0$ are invertible.}

\noindent PROOF: The proof for the expansions in (6.13) is
similar to the proof of Proposition~4.4 of [4]. The invertibility
of $\Cal D_0$ follows from the fact that it consists of invertible
blocks and is given by
$$\Cal D_0=\text{diag}\{I_{\nu-\mu},J_{n_\mu+1}(\lambda_{n_\mu+1}),
\dots,J_{n_\kappa}(\lambda_{n_\kappa})\},$$
where $I_{\nu-\mu}$ is the identity matrix
of size $(\nu-\mu)$ with
$\mu$ and $\nu$ denoting the respective geometric and algebraic
multiplicities of the zero eigenvalue of
$J(0),$ and the $J_{n_\alpha}(\lambda_\alpha)$
are the Jordan block matrices appearing in (6.7)
corresponding to the nonzero eigenvalues for
$\alpha=\mu+1,\dots,\kappa.$
 From (6.1) and (6.10), as the $(s,j)$-entry of the matrix
$\Cal A_1$ we get
$$(\Cal A_1)_{sj}=-i v_{s n_s}^\dagger R u_{j1},$$
where $R$ is the matrix appearing in (6.1) and (6.2). By (5.63)
and Proposition~5.7 we know that $R$ acts as an invertible map
from  $\text{Ker}\, [J(0)]$ to $\text{Ker}\, [J(0)^\dagger].$
Recall from (6.3) and (6.5) that $\{u_{11},u_{21},\dots,u_{\mu
j}\}$ is the Jordan basis for $\text{Ker}\, [J(0)]$ and $\{v_{1
n_1},v_{2 n_2},\dots,v_{\mu n_\mu}\}$ is the Jordan basis for
$\text{Ker}\, [J(0)\dagger].$ Thus, the matrix $\Cal A_1$ is
nothing but, apart from the factor $(-i),$ the matrix
representation of the invertible map $R$ with respect to the
Jordan basis and the adjoint Jordan basis. Thus, $\Cal A_1$ is
invertible. \qed

\noindent {\bf Theorem 6.2} {\it Assume that $V$ in (1.1)
is selfadjoint
and belongs to $L^1_1(\bR^+).$ Then, the asymptotics
as $k\to 0$ in $\bCpb$ of the inverse of the
matrix $\Cal Z(k)$ appearing in (6.12)
is given by}
$$\Cal Z(k)^{-1}=\bm \ds\frac{1}{k}\,
\Cal A_1^{-1}[I_\mu+o(1)]&-\Cal A_1^{-1}\Cal B_1 \Cal D_0^{-1}+o(1)\\
\stretch
-\Cal D_0^{-1} \Cal C_1 \Cal A_1^{-1}+o(1)&\Cal D_0^{-1}+O(k)\endbm,
\tag 6.14$$
{\it where $\Cal A_1,$ $\Cal C_1,$ and $\Cal D_0$
are the constant matrices appearing in (6.13) and
the invertibility of $\Cal A_1$ and $\Cal D_0$
is assured in Theorem~6.1.}

\noindent PROOF: The proof is exactly the same as the proof
of Proposition~4.5(i) of [4]. We will use the decomposition formula [4,10]
$$\bm I_\mu&-\Cal B\Cal D^{-1}\\
\stretch
0&I_{n-\mu}\endbm
\bm \Cal A&\Cal B\\
\stretch
\Cal C& \Cal D\endbm\bm
I_\mu&0\\
\stretch
-\Cal D^{-1}\Cal C&I_{n-\mu}\endbm
=\bm \Cal A-\Cal B\Cal D^{-1}\Cal C&0\\
\stretch
0&\Cal D\endbm.
\tag 6.15$$
Thus, as seen from (6.15)
for the matrix $\Cal Z(k)$ defined
in (6.12) we have
$$\Cal Z^{-1}=\bm \Cal A&\Cal B\\
\stretch
\Cal C& \Cal D\endbm^{-1}=
\bm
I_\mu&0\\
\stretch
-\Cal D^{-1}\Cal C&I_{n-\mu}\endbm
\bm (\Cal A-\Cal B\Cal D^{-1}\Cal C)^{-1}&0\\
\stretch
0&\Cal D^{-1}\endbm
\bm I_\mu&-\Cal B\Cal D^{-1}\\
\stretch
0&I_{n-\mu}\endbm,
$$
or equivalently
$$\Cal Z(k)^{-1}=
\bm
(\Cal A-\Cal B\Cal D^{-1}\Cal C)^{-1}&-(\Cal A-\Cal B\Cal D^{-1}\Cal C)^{-1}
\Cal B\Cal D^{-1}
\\
\stretch
-\Cal D^{-1}\Cal C(\Cal A-\Cal B\Cal D^{-1}\Cal C)^{-1}&
\Cal D^{-1}\Cal C(\Cal A-\Cal B\Cal D^{-1}\Cal C)^{-1}\Cal B\Cal D^{-1}+\Cal D^{-1}\endbm.\tag 6.16
$$
Finally, using (6.13) in (6.16) and the fact that $\Cal A_1$
and $\Cal D_0$ are invertible, we get (6.14). Let us note that
we have written $(1/k)\,\Cal A_1^{-1}[I_\mu+o(1)]$ in the top
left block in (6.14) whereas that term was written as
$(1/k)\,\Cal A_1^{-1}+o(1/k)$ in [4]. The two expressions are
certainly equivalent because we can always premultiply the
$o(1/k)$-term by $\Cal A_1^{-1} \Cal A_1.$ \qed

We are now ready to evaluate the small-$k$ limit of
the Jost matrix $J(k)$ defined in (4.3), its inverse $J(k)^{-1},$ and
the scattering matrix $S(k)$ defined in (4.6). From (6.2) and
(6.12) we see that
the Jost matrix $J(k)$
is given by
$$J(k)=f(-k^*,a)^\dagger\,[f(0,a)^\dagger]^{-1}
\Cal S P_2^{-1} \Cal Z(k)\,P_1^{-1} \Cal S^{-1},\tag 6.17$$
where the small-$k$ limit will be evaluated with the help
of (3.8), (6.12), and (6.13). On the other hand, from  (6.17)
we get
$$J(k)^{-1}=
\Cal S \,P_1\,\Cal Z(k)^{-1}P_2  \Cal S^{-1}f(0,a)^\dagger
\,[f(-k^*,a)^\dagger]^{-1},\tag 6.18$$
where the small-$k$ limit will be evaluated with the help
of (3.8) and (6.14).
Thus, using (6.17) and (6.18) in (4.6) we obtain
$$S(k)=-f(k,a)^\dagger\,[f(0,a)^\dagger]^{-1}
\Cal S P_2^{-1} \Cal Z(-k)\,\Cal Z(k)^{-1}P_2  \Cal S^{-1}f(0,a)^\dagger
\,[f(-k,a)^\dagger]^{-1}.\tag 6.19$$
In evaluating the small-$k$ limits of
$J(k),$ $J(k)^{-1},$ and $S(k),$
we will use a consequence of (3.8), namely
for $k\to 0$ in $\bCpb$ we have
$$f(0,a)^\dagger\,[f(-k^*,a)^\dagger]^{-1}=I_n+o(1),\quad
f(k,a)^\dagger[f(0,a)^\dagger]^{-1}=I_n+o(1).\tag 6.20$$

\noindent {\bf Theorem 6.3} {\it Assume that $V$ in (1.1)
is selfadjoint
and belongs to $L^1_1(\bR^+).$ Then, as $k\to 0$ in $\bCpb$ the Jost matrix
$J(k)$ has the behavior
$$J(k)=\Cal S P_2^{-1}
\bm k \Cal A_1+o(k)&k \Cal B_1 \Cal A_1+o(k)\\
\stretch
k \Cal C_1+o(k)&\Cal D_0+o(1)\endbm P_1 \Cal S^{-1},\tag 6.21$$
the inverse
Jost matrix
$J(k)^{-1}$ has the behavior, as $k\to 0$ in $\bCpb,$
$$J(k)^{-1}=\Cal S P_1
\bm \ds\frac{1}{k}\, \Cal A_1^{-1}[I_\mu+o(1)]&
- \Cal A_1^{-1}\Cal B_1  \Cal D_0^{-1}+o(1)\\
\stretch
-\Cal D_0^{-1}\Cal C_1\Cal A_1^{-1}+o(1)&
\Cal D_0^{-1}+o(1)\endbm P_2\Cal S^{-1},\tag 6.22$$
and the scattering matrix
$S(k)$ defined in (4.6) is continuous at $k=0$  and
we have $S(k)=S(0)+o(1)$ as $k\to 0$ in $\bR$ with}
$$S(0)=\Cal S P_2^{-1}
\bm I_\mu& 0\\
\stretch
2\Cal C_1\Cal A_1^{-1}&-I_{n-\mu}\endbm
P_2  \Cal S^{-1},\tag 6.23$$
{\it where $\Cal A_1,$ $\Cal B_1,$ $\Cal C_1,$ and $\Cal D_0$
are the matrices appearing in (6.13),
$\mu$ is the geometric multiplicity
of the zero eigenvalue of the zero-energy
 Jost matrix $J(0),$ $P_1$ and
 $P_2$ are the permutation operators appearing in (6.11),
 and $\Cal S$ is the matrix appearing in (6.7).}

\noindent PROOF: Using (6.13) and (6.20) in (6.17) we get
(6.21). Using (6.14) and (6.20) in (6.18) we obtain (6.22).
Finally, using (6.12)-(6.14) we obtain
$$\Cal Z(-k)\,\Cal Z(k)^{-1}=\bm -I_\mu+o(1)& O(k)\\
\stretch -2\Cal C_1\Cal A_1^{-1}+o(1)&I_{n-\mu}+O(k)\endbm,\qquad k\to 0\text{ in }\bR.
\tag 6.24$$
Then, using (6.24) and (6.20)
in (6.19) we get (6.23) and
$S(k)=S(0)+o(1)$ as $k\to 0$ in $\bR.$ \qed

\vskip 10 pt
\noindent {\bf 7. EXAMPLES}
\vskip 3 pt

In this section, we will check the validity of our formula (6.23) for some selfadjoint boundary conditions.

\noindent {\bf Example 7.1 (The $\delta'$ boundary condition)} As our first example, let us use the $3\times 3$ versions of
$A$ and $B$ from p. S116 of [25], with $A$ and $B$ satisfying (1.11)-(1.13)
and given by
$$A=\bm 1&0&-a\\
\stretch
-1&1&0\\
\stretch
0&-1&0\endbm,\quad
B=\bm 0&0&-1\\
\stretch
0&0&-1\\
\stretch
0&0&-1\endbm,$$
where $a$ is a real parameter and the potential $V$ is zero. From (4.3) with
$f(k,x)=e^{ikx}I_3 ,$ we obtain $J(k),$ and from (4.6) we obtain the scattering matrix
$S(k),$ and we have
$$J(k)=
\bm -ik&0&-1+iak\\
\stretch
ik&-ik&-1\\
\stretch
0&ik&-1\endbm,\quad
S(k)=\bm \ds\frac{i+ak}{3i+ak}&\ds\frac{-2i}{3i+ak}&\ds\frac{-2i}{3i+ak}\\
\stretch
\ds\frac{-2i}{3i+ak}&\ds\frac{i+ak}{3i+ak}&\ds\frac{-2i}{3i+ak}\\
\stretch
\ds\frac{-2i}{3i+ak}&\ds\frac{-2i}{3i+ak}&\ds\frac{i+ak}{3i+ak}\endbm.
\tag 7.1
$$
 From (7.1) we see that $S(k)$ is continuous at $k=0$ and
$$S(0)=\bm \ds\frac13&-\ds\frac{2}{3}&-\ds\frac{2}{3}\\
\stretch
-\ds\frac{2}{3}&\ds\frac13&-\ds\frac{2}{3}\\
\stretch
-\ds\frac{2}{3}&-\ds\frac{2}{3}&\ds\frac13\endbm.
\tag 7.2
$$
On the other hand, we see that $J(0)$ is given by
$$J(0)=\bm 0&0&-1\\
\stretch
0&0&-1\\
\stretch
0&0&-1\endbm,$$
and hence its eigenvalues are $\lambda_1=0,$ $\lambda_2=0,$
and $\lambda_3=-1,$ with respective eigenvectors
$$u_{11}=\bm 1\\
\stretch
0\\
\stretch
0\endbm,\quad u_{21}=
\bm 0\\
\stretch
1\\
\stretch
0\endbm,\quad
u_{31}=
\bm 1\\
\stretch
1\\
\stretch
1\endbm.$$
Thus, in the notation of Section~6 we have
$n_1=1,$ $n_2=1,$ and $n_3=1,$ with $\kappa=3.$ Our permutation
operators $P_1$ and $P_2$ are given by
$P_1=I_3$ and $P_2=I_3,$ and the matrix $\Cal S$ appearing in (6.7)
and
the matrix $\Cal Z(k)$ appearing in (6.12) are given by
$$\Cal S= \bm 1&0&1\\
\stretch
0&1&1\\
\stretch
0&0&1\endbm,\quad
\Cal Z(k)=\bm -ik&-ik&(a-2)ik\\
\stretch
ik&-2ik&-ik\\
\stretch
0&ik&-1+ik\endbm.\tag 7.3$$
 From (7.3), with the help of (6.13) we get
$$\Cal A_1=
\bm -i&-i\\
\stretch
i&-2i\endbm,\quad
\Cal B_1=
\bm (a-2)i\\
\stretch
-i\endbm,\quad
\Cal C_1=
\bm 0&i
\endbm,\quad
\Cal D_0=\bm -1\endbm.
$$
Since $\mu=\nu=2$ and $n=3$ in our example, we evaluate (6.23)
and confirm that the right hand side in (6.23) coincides with
the matrix in (7.2).

\noindent {\bf Example 7.2 (The Kirchhoff boundary condition)} The procedure in this example is similar to
that of Example~7.1. We use the $3\times 3$ versions of
$A$ and $B$ from p. S117 of [25]. Using $V=0$ and
$$A=\bm 0&0&1\\
\stretch
0&0&1\\
\stretch
0&0&1\endbm,\quad
B=\bm -1&0&0\\
\stretch
1&-1&0\\
\stretch
0&1&0\endbm,$$
we obtain
$$J(k)=
\bm -1&0&-ik\\
\stretch
1&-1&-ik\\
\stretch
0&1&-ik\endbm,\quad
S(k)=S(0)=\bm -\ds\frac{1}3&\ds\frac{2}{3}&\ds\frac{2}{3}\\
\stretch
\ds\frac{2}{3}&-\ds\frac{1}3&\ds\frac{2}{3}\\
\stretch
\ds\frac{2}{3}&\ds\frac{2}{3}&-\ds\frac{1}3\endbm.
\tag 7.4
$$
On the other hand, we have
$J(0)$ given by
$$J(0)=\bm -1&0&0\\
\stretch
1&-1&0\\
\stretch
0&1&0\endbm.$$
The eigenvalues of $J(0)$ are $\lambda_1=0$ with $n_1=1$
and $\lambda_2=-1$ with $n_2=2.$ We further have
$$\mu=\nu=1,\quad n=3,\quad P_1=P_2=I_3,\tag 7.5$$
$$\Cal S= \bm 0&0&-1/\sqrt{2}\\
\stretch
0&-1/\sqrt{2}&1/\sqrt{2}\\
\stretch
1&1/\sqrt{2}&0\endbm,\quad
\Cal Z(k)=\bm -3ik&-3ik/\sqrt{2}&0\\
\stretch
2\sqrt{2}ik&-1+2ik&1\\
\stretch
\sqrt{2}ik&ik&-1\endbm,\tag 7.6$$
$$\Cal A_1=
\bm -3i\endbm,\quad
\Cal B_1=
\bm -3i/\sqrt{2}&0\\
\endbm,\quad
\Cal C_1=
\bm 2\sqrt{2}i\\
\stretch
\sqrt{2}i
\endbm,\quad
\Cal D_0=\bm -1&1\\
\stretch
0&-1\endbm.
\tag 7.7$$
Using the information in (7.5)-(7.7) in (6.23) we can verify
that the right side in (6.23) coincides
with $S(0)$ given in (7.4).

\noindent {\bf Example 7.3 (The XOR gate boundary condition)}
Recall that the XOR gate is a digital logic gate implementing
an exclusive disjunction, yielding false if the two inputs
agree and yielding true if the inputs disagree. The procedure
in this example is similar to that of Example~7.1. We use the
$4\times 4$ versions of the matrices $A$ and $B$ from [24].
Using $V=0$ and
$$A=\bm i/a&0&0&0\\
\stretch
0&i/a&0&i/(2a)\\
\stretch
0&0&i/(2a)&i/(2a)\\
\stretch
0&0&i/(2a)&i/(2a)
\endbm,\quad
B=\bm 0&0&0&0\\
\stretch
0&0&0&0\\
\stretch
0&0&-1/2&1/2
\\
\stretch
0&0&1/2&-1/2
\endbm,$$
where $a$ is a real parameter,
we obtain
$$J(k)=
\bm k/a&0&0&0\\
\stretch
0&k/a&0&k/(2a)\\
\stretch
0&0&(k-a)/(2a)&(k+a)/(2a)
\\
\stretch
0&0&(k+a)/(2a)&(k-a)/(2a)\endbm,\quad
S(k)=S(0)=\bm 1&0&0&0\\
\stretch
0&1&0&0\\
\stretch
0&0&0&1
\\
\stretch
0&0&1&0\endbm.
\tag 7.8
$$
On the other hand, we have
$J(0)$ given by
$$J(0)=\bm 0&0&0&0\\
\stretch
0&0&0&0\\
\stretch
0&0&-1/2&1/2
\\
\stretch
0&0&1/2&-1/2
\endbm.$$
The eigenvalues of $J(0)$ are $\lambda_1=0$ with $n_1=1,$
$\lambda_2=0$ with $n_2=1,$ $\lambda_3=0$ with $n_3=1,$
and $\lambda_4=-1$ with $n_4=1.$ We further have
$$\mu=\nu=3,\quad n=4,\quad
P_1=P_2=I_4,\tag 7.9$$
$$\Cal S= \bm 1&0&0&0\\
\stretch
0&1&0&0\\
\stretch
0&0&1&-1
\\
\stretch
0&0&1&1
\endbm,\quad
\Cal Z(k)=\bm k/a&0&0&0\\
\stretch
0&k/a&k/(2a)&k/(2a)\\
\stretch
0&0&k/a&0
\\
\stretch
0&0&0&-1
\endbm,\tag 7.10$$
$$\Cal A_1=
\bm 1/a&0&0
\\
\stretch
0&1/a&1/(2a)
\\
\stretch
0&0&1/a
\endbm,\quad
\Cal B_1=
\bm 0\\
\stretch
1/(2a)
\\
\stretch
0
\endbm,\quad
\Cal C_1=
\bm 0&0&0
\endbm,\quad
\Cal D_0=\bm -1\endbm.
\tag 7.11$$
Using the information in (7.9)-(7.11) in (6.23) we can verify
that the right side in (6.23) coincides
with $S(0)$ given in (7.8).

\noindent {\bf Example 7.4} The procedure in this example is again
similar to
that of Example~7.1. We use $V=0,$ and for some real parameters $a$
and $b,$
we let $A$ and $B$ be
$$A=\bm 2&1&a\\
\stretch
0&0&b\\
\stretch
1&1&c
\endbm,\quad
B=\bm 0&0&0\\
\stretch
0&0&1\\
\stretch
0&0&0
\endbm.$$
We obtain
$$J(k)=
\bm -2ik&-ik&-iak\\
\stretch
0&0&1-ibk\\
\stretch
-ik&-ik&-ick\endbm,\quad
S(k)=\bm 1&0&0\\
\stretch
0&\ds\frac{-i+bk}{i+bk}&0\\
\stretch
0&0&1
\endbm,
$$
and hence
$$S(0)=\bm 1&0&0\\
\stretch
0&-1&0\\
\stretch
0&0&1
\endbm.\tag 7.12$$
On the other hand, the matrix
$J(0),$ which given by
$$J(0)=\bm 0&0&0\\
\stretch
0&0&1\\
\stretch
0&0&0
\endbm,$$
has eigenvalues $\lambda_1=0$ with $n_1=1$ and
$\lambda_2=0$ with $n_2=2.$ We further have
$$\mu=2,\quad\nu=3,\quad n=3,\quad
P_1=I_3,\quad P_2=\bm 1&0&0\\
\stretch
0&0&1\\
\stretch
0&1&0
\endbm,\tag 7.13$$
$$\Cal S= \bm 1&0&0\\
\stretch
0&1&0\\
\stretch
0&0&1
\endbm,\quad
\Cal Z(k)=\bm -2ik&-ik&-iak\\
\stretch
-ik&-ik&-ick\\
\stretch
0&0&1-ibk
\endbm,\tag 7.14$$
$$\Cal A_1=
\bm -2i&-i
\\
\stretch
-i&-i
\endbm,\quad
\Cal B_1=
\bm -ia\\
\stretch
-ic
\endbm,\quad
\Cal C_1=
\bm 0&0
\endbm,\quad
\Cal D_0=\bm 1\endbm.
\tag 7.15$$
Using the information in (7.13)-(7.15) in (6.23) we can verify
that the right hand side in (6.23) coincides
with $S(0)$ given in (7.12).

\vskip 10 pt

\noindent {\bf Acknowledgment.} The research leading to this
article was supported in part by Consejo Nacional de Ciencia y
Tecnolog\'{\i}a (CONACYT) under project CB2008-99100-F, the
Texas Norman Hackerman Advanced Research Program (NHARP) under
grant no. 003656-0046-2007, and by the Department of Defense
under grant number DOD-BC063989.

\vskip 10 pt

\noindent {\bf{References}}

\vskip 3 pt

\item{[1]} Z. S. Agranovich and V. A. Marchenko, {\it The inverse problem of
scattering theory,} Gordon and Breach, New York, 1963.

\item{[2]} T. Aktosun,
{\it Factorization and small-energy asymptotics for the radial Schr\"odinger equation,}
J. Math. Phys. {\bf 41}, 4262--4270 (2000).

\item{[3]} T. Aktosun and M. Klaus,
{\it Small-energy asymptotics for the Schr\"odinger equation on the line,}
Inverse Problems {\bf 17}, 619--632 (2001).

\item{[4]} T. Aktosun, M. Klaus, and C. van der Mee,
{\it Small-energy asymptotics of the scattering matrix for the matrix
Schr\"odinger equation on the line,}
J. Math. Phys. {\bf 42}, 4627--4652 (2001).

\item{[5]} T. Aktosun and R. Weder, {\it
Inverse spectral-scattering problem with two sets of discrete
spectra for the radial Schr\"odinger equation,} Inverse Problems {\bf 22}, 89--114
(2006).

\item{[6]} G. Berkolaiko, R. Carlson, S. A. Fulling, and P. Kuchment (eds.),
{\it Quantum graphs and their applications,} Contemporary Mathematics, 415,
Amer. Math. Soc., Providence, RI, 2006.

\item{[7]} J. Boman and P. Kurasov,
{\it Symmetries of quantum graphs and the inverse scattering problem,}
Adv. Appl. Math. {\bf 35}, 58--70 (2005).

\item{[8]} G. Borg,
 {\it Uniqueness theorems in the spectral theory of $y''+(\lambda-q(x))\,y=0,$}
Proc. 11th Scandinavian Congress of Mathematicians, Johan
Grundt Tanums Forlag, Oslo, pp. 276--287, 1952.

\item{[9]} P. Deift and E. Trubowitz, {\it Inverse scattering
on the line,} Commun. Pure Appl. Math. {\bf 32}, 121--251 (1979).

\item{[10]} H. Dym,
{\it Linear algebra in action,} Amer. Math. Soc., Providence, R.I.,
2007.

\item{[11]} P. Exner, J. P. Keating, P. Kuchment, T. Sunada, and A. Teplyaev (eds.),
{\it Analysis on graphs and its applications,}
Proc. Symposia in Pure Mathematics, 77,
Amer. Math. Soc., Providence, RI, 2008.

\item{[12]} L. D. Faddeev, {\it
Properties of the $S$-matrix of the one-dimensional Schr\"odinger equation,} Amer. Math. Soc. Transl. {\bf 65} (ser. 2), 139--166 (1967).

\item{[13]} I. M. Gel'fand and B. M. Levitan,
{\it On the determination of a differential equation from its spectral function,} Amer. Math. Soc. Transl. {\bf 1} (ser. 2), 253--304 (1955).

\item{[14]} N. I. Gerasimenko,
{\it The inverse scattering problem on a noncompact graph,}
Theoret. Math. Phys. {\bf 75}, 460--470 (1988).

\item{[15]} N. I. Gerasimenko and B. S. Pavlov,
{\it A scattering problem on noncompact graphs,}
Theoret. Math. Phys. {\bf 74}, 230--240 (1988).

\item{[16]} F. Gesztesy and B. Simon,
{\it Uniqueness theorems in inverse spectral theory for one-dimensional
Schr\"odinger operators,} Transact. Amer. Math. Soc. {\bf 348}, 349--373 (1996).

\item{[17]} B. Gutkin and U. Smilansky,
{\it Can one hear the shape of a graph?}
J. Phys. A {\bf 34}, 6061--6068 (2001).

\item{[18]} M. S. Harmer, {\it Inverse scattering for the matrix Schr\"odinger
operator and Schr\"odinger operator on
graphs with general self-adjoint boundary conditions,}
ANZIAM J. {\bf 44}, 161--168 (2002).

\item{[19]} M. S. Harmer, {\it The matrix Schr\"odinger operator
and Schr\"odinger operator on graphs,} Ph.D. thesis, University of
Auckland, New Zealand, 2004.

\item{[20]} M. Harmer, {\it Inverse scattering on matrices with boundary conditions,}
J. Phys. A {\bf 38}, 4875--4885 (2005).

\item{[21]} M. Klaus, {\it
Low-energy behaviour of the scattering matrix for the Schr\"odinger equation on the line,} Inverse Problems {\bf 4}, 505--512 (1988).

\item{[22]} M. Klaus, {\it Exact behavior of Jost functions at low energy,}
J. Math. Phys. {\bf 29}, 148--154 (1988).

\item{[23]} V. Kostrykin and R. Schrader,
{\it Kirchhoff's rule for quantum wires,} J. Phys. A {\bf 32}, 595--630
(1999).

\item{[24]} V. Kostrykin and R. Schrader,
{\it Kirchhoff's rule for quantum wires. II: The inverse problem with possible applications to quantum computers,} Fortschr. Phys. {\bf 48}, 703--716
(2000).

\item{[25]} P. Kuchment,
{\it Quantum graphs. I. Some basic structures,}
Waves Random Media {\bf 14}, S107--S128 (2004).

\item{[26]} P. Kuchment,
{\it Quantum graphs. II. Some spectral properties of quantum and combinatorial graphs,}
J. Phys. A {\bf 38}, 4887--4900 (2005).

\item{[27]} P. Kurasov and M. Nowaczyk,
{\it Inverse spectral problem for quantum graphs,}
J. Phys. A {\bf 38}, 4901--4915 (2005).

\item{[28]} P. Kurasov and F. Stenberg,
{\it On the inverse scattering problem on branching graphs,}
J. Phys. A {\bf 35}, 101--121 (2002).

\item{[29]} B. M. Levitan, {\it
Inverse Sturm Liouville Problems,} VNU Science Press, Utrecht, 1987.

\item{[30]} V. A.
Marchenko, {\it Some questions in the theory of one-dimensional
linear differential operators of the
second order,} Amer. Math. Soc. Transl. {\bf 101} (ser. 2), 13--104 (1973).

\item{[31]} V. A. Marchenko,
{\it Sturm-Liouville operators and applications,}
Birkh\"auser, Basel, 1986.

\end